%% file: ccs.tex
\documentclass[sigconf]{acmart}

\input{header}

  \providecommand\BibTeX{{%
    \normalfont B\kern-0.5em{\scshape i\kern-0.25em b}\kern-0.8em\TeX}}

\copyrightyear{2023} 
\acmYear{2023} 
\setcopyright{acmlicensed}\acmConference[CCS '23]{Proceedings of the 2023 ACM SIGSAC Conference on Computer and Communications Security}{November 26--30, 2023}{Copenhagen, Denmark}
\acmBooktitle{Proceedings of the 2023 ACM SIGSAC Conference on Computer and Communications Security (CCS '23), November 26--30, 2023, Copenhagen, Denmark}
\acmPrice{15.00}
\acmDOI{10.1145/3576915.3623076}
\acmISBN{979-8-4007-0050-7/23/11}

\title{\salsys{}: 
 A Machine Learning Attack On LWE with Binary Secrets}

 \author{Cathy Yuanchen Li}
 \email{cathyli@uchicago.edu}
 \affiliation{%
  \institution{Meta AI}
  \city{Seattle}
  \country{USA}}
 \authornote{Co-first authors.}

\author{Jana Sot\'{a}kov\'{a}}
 \email{ja.sotakova@gmail.com}
 \affiliation{%
  \institution{Meta AI}
  \city{Seattle}
  \country{USA}}
\authornotemark[1]

\author{Emily Wenger}
\email{ewenger@uchicago.edu}
 \affiliation{%
  \institution{The University of Chicago}
  \city{Chicago}
  \country{USA}}

\author{Mohamed Malhou}
\email{mohamed.malhou@polytechnique.edu}
 \affiliation{%
  \institution{Meta AI}
  \city{Paris}
  \country{France}}

\author{Evrard Garcelon}
\email{evrard.garcelon@gmail.com}
 \affiliation{%
  \institution{Meta AI}
  \city{Paris}
  \country{France}}

\author{Fran\c{c}ois Charton}
\email{fcharton@meta.com}
 \affiliation{%
  \institution{Meta AI}
  \city{Paris}
  \country{France}}
\authornote{Co-senior authors, corresponding author: \texttt{fcharton@meta.com}}

\author{Kristin Lauter}
\email{klauter@meta.com}
 \affiliation{%
  \institution{Meta AI}
  \city{Seattle}
  \country{USA}}
\authornotemark[2]

\begin{document}

 \renewcommand{\shortauthors}{Cathy Yuanchen Li et al.}

\begin{abstract}

{\it Learning With Errors} (LWE) is a hard math problem underpinning many proposed post-quantum cryptographic (PQC) systems. The only PQC {\it Key Exchange Mechanism} (KEM) standardized by NIST \cite{nist2022finalists}
is based on module~LWE, and current publicly available PQ Homomorphic Encryption (HE) libraries are based on ring~LWE \cite{HES}. The security of LWE-based PQ cryptosystems is critical, but certain implementation choices could weaken them. One such choice is sparse binary secrets, desirable for PQ HE schemes for efficiency reasons.
Prior work \salsa{}~\cite{wengersalsa} demonstrated a machine learning-based attack on 
LWE with sparse binary secrets in small dimensions ($n \le 128$) and low Hamming weights ($h \le 4$). However, this attack assumes access to millions of eavesdropped LWE samples and fails at higher Hamming weights or dimensions.

We present \system{}, an enhanced machine learning attack on LWE with sparse binary secrets, which recovers secrets in much larger dimensions (up to $n=350$) and with larger Hamming weights (roughly $n/10$, and up to $h=60$ for $n=350$). We achieve this dramatic improvement via a novel {\it preprocessing step}, which allows us to generate training data from a linear number of eavesdropped LWE samples ($4n$) and changes the distribution of the data to improve transformer training. We also improve the secret recovery methods of \salsa{} and introduce a novel cross-attention recovery mechanism allowing us to read off the secret directly from the trained models. While \system{} does not threaten NIST's proposed LWE standards, it demonstrates significant improvement over \salsa{} and could scale further, highlighting the need for future investigation into machine learning attacks on LWE with sparse binary secrets.

\end{abstract}



\maketitle

\input{intro}

\input{back}

\input{method}
\input{attack}

\input{results}

\input{ablation}
\input{comparison}

\input{discussion}

\bibliographystyle{acm}
\balance
{\footnotesize 
\bibliography{references}}

\appendix
\input{appendix}
\end{document}

%% file: header.tex
\usepackage{microtype}
\usepackage{graphicx}
\usepackage{subfigure}
\usepackage{booktabs} 
\usepackage{multirow}
\usepackage{balance}

\usepackage{hyperref}
\usepackage{caption}

\usepackage{amsmath,amsthm,amscd,url,enumerate, mathtools}
\usepackage[utf8]{inputenc}
\usepackage{wrapfig}
\usepackage{pdflscape}
\usepackage{tikz}
\usetikzlibrary{decorations.markings,arrows}
\usetikzlibrary{positioning,arrows,fit}

\usepackage{tikz-cd}
\usepackage{amsfonts}
\usepackage{graphicx}
\usepackage{mathrsfs}
\usepackage{dsfont}
\usepackage{enumitem}
\usepackage{array}
\usepackage{verbatim}
\usepackage{soul}

\newcolumntype{?}{!{\vrule width 1pt}}
\usepackage[capitalize,noabbrev]{cleveref}

\newcommand{\para}[1]{{\vspace{2pt} \bf \noindent #1}}  
\newenvironment{packed_itemize}{
\begin{list}{\labelitemi}{\leftmargin=1em}
\setlength{\itemsep}{1pt}                                                           
\setlength{\parskip}{0pt}                                                                                 \setlength{\parsep}{0pt}                                                                                  \setlength{\headsep}{0pt}                                                                                 \setlength{\topskip}{0pt}                                                                                 \setlength{\topmargin}{0pt}                                                                               \setlength{\topsep}{0pt}                                                                                  \setlength{\partopsep}{0pt}                                                                               }{\end{list}}                                                                                                                                  

\newcommand{\ZZ}{\mathbb{Z}}

\DeclarePairedDelimiter\ceil{\lceil}{\rceil}

\newcommand{\T}{\mathcal{M}}
\newcommand{\system}{{\textsc{Picante}}}
\newcommand{\salsys}{{\textsc{Salsa} \textsc{Picante}}}
\newcommand{\salsa}{{\textsc{Salsa}}}
\newcommand{\tinyLWE}{{\textsc{TinyLWE}}}
\newcommand{\lwe}{\textsc{LWE}}
\newcommand{\fplll}{{\textit{fplll}}}

\newcommand{\rev}[1]{#1} 

%% file: intro.tex
\section{Introduction}
\label{sec:intro}

The race for post-quantum cryptography (PQC) is well underway. A large-scale quantum computer could solve the hard math problems underpinning most deployed public-key cryptographic systems, like RSA~\cite{rivest1978method}, in polynomial time. Small-scale quantum computers have already been built. Consequently, new post-quantum cryptographic systems were proposed and considered for standardization by US National Institute of Standards and Technology (NIST) in the 5-year PQC competition. In July 2022, NIST standardized 4 schemes from the PQC competition~\cite{nist2022finalists}. The only key encapsulation mechanism selected\textemdash CRYSTALS-Kyber~\cite{crys_kyber}\textemdash and one of the three signature schemes\textemdash CRYSTALS-Dilithium~\cite{crys_dilithium}\textemdash are based on the mathematical hardness assumption known as {Learning With Errors} (LWE)~\cite{Reg05}. LWE is also used in proposed PQ homomorphic encryption schemes~\cite{HES}.

LWE works as follows: given an integer modulus $q$, a dimension $n$, and a secret vector ${\bf s} \in \mathbb{Z}^n_q$, the {\it Learning With Errors problem} is to recover ${\bf s}$ given many random vectors and their noisy inner products with ${\bf s}$. These noisy inner products are computed by taking random vector ${\bf a} \in \mathbb{Z}^n_q$ and producing $b := \textbf{a} \cdot \textbf{s}+e \mod q$, where $e$ is an ``error'' term sampled from a narrow discrete Gaussian distribution (i.e. taking small values). 
The adversary is then given the {\it samples} $(\mathbf{a}, b)$ and attempts to use these to recover ${\bf s}$.

The basic LWE problem is assumed to be hard for both classical and quantum adversaries~\cite{Pei09a,LM09_hardness_lwe_exp,BLPRS13_poly_modulus_hardness,regev:quantum,Reg05}. Variants of LWE, like module-LWE or ring-LWE\textemdash on which the NIST-standards and HE schemes are based\textemdash add structure to the basic LWE problem, making them potentially easier than LWE. Classical attacks on LWE and its variants typically rely on algebraic techniques for lattice reduction to recover the secret $\bf{s}$ from pairs $({\bf a},b)$~\cite{LLL, CN11_BKZ}. 
The error $e$ added to ${\bf a} \cdot {\bf s}$ to compute $b$ adds noise, making algebraic solutions difficult.
Fundamentally, the LWE hardness assumption is that it is hard to learn from noisy data, rendering LWE secret recovery computationally expensive. 

On the other hand, the whole field of machine learning (ML) depends on the fact that it is possible to train machines to learn from noisy data. Recent advances in model architectures (e.g.~\cite{transformer17}) and training techniques have allowed ML models to glean meaningful trends even from noisy unstructured data. Although LWE samples ($\bf a$, $b$) are noisy, they are highly structured, a fact that ML models can exploit for learning. Thus, it is worthwhile to investigate whether ML-based attacks can enable LWE secret-recovery.

Prior work, \salsa{} \cite{wengersalsa}, provided an initial proof-of-concept for ML-based LWE attacks. \salsa{} demonstrated the feasibility of recovering sparse binary secrets, attractive for example in HE applications, for LWE problems with relatively small parameters. Given many LWE samples, \salsa{} trains ML models to learn the underlying structure of the LWE problem, then leverages the trained model to recover the LWE secret.  If the models learn to predict $b$ from ${\bf a}$ (even with low accuracy), \salsa{} can recover the secret ${\bf s}$.

Although promising, \salsa{} has significant limitations. For the largest dimension $n=128$, \salsa{} can only recover Hamming weight $h \leq 3$. In comparison, real-world LWE schemes with binary secrets (for homomorphic encryption) start at dimension $n=512$ or $n=1024$. 
\salsa{} also requires millions of LWE samples $({\bf a},b)$ for model training, but a real-world attacker would likely only have access to a few samples. Making \salsa{}'s approach realistic requires scaling up the parameters of solvable LWE problems~($n,h$, and modulus $q$) while reducing the number of required samples. 

\para{Contributions.} In this work, we propose \system{}, an enhanced ML-based attack on the LWE hardness assumption. \system{} leverages basic principles of the original \salsa{} attack~\cite{wengersalsa}\textemdash transformer training, secret recovery\textemdash while introducing several novel techniques. This enhanced attack allows recovery of high-dimensional binary secrets with Hamming weight roughly $n/10$ or beyond, requiring only $4n$ samples for training. Table~\ref{tab:best_results} shows the largest Hamming weights we recover for each dimension.

\begin{table}[h!]
    \centering
    \small 
    \begin{tabular}{lcccccc} \toprule
 {Dimension}        &  80 & 150 & 200 & 256 & 300 & 350 \\ \midrule 
 log $q$  &   7 &  13 & 17 &  23 & 27 & 32\\ 
    {highest $h$}     &  9 &  13 & 22 & 31 & 33 & 60 \\
    \midrule
    \rev{$\#$ possible secrets} &   \rev{$2^{32}$} &  \rev{$2^{61}$} &  \rev{$2^{97}$} &  \rev{$2^{133}$}  &  \rev{$2^{147}$} & \rev{$2^{227}$} \\ \bottomrule
    \end{tabular}
    \vspace{0.1cm}
    \caption{\small\textbf{\salsys{}'s highest recovered secret Hamming weights~$h$}. \rev{The bottom row lists the approximate number of possible secrets for each $n$/$h$ combination, for comparison with brute force guessing attacks.}}
    \vspace{-0.4cm}
    \label{tab:best_results}
\end{table}

As in \salsa{} \cite[Table 4]{wengersalsa}, we observe that it is easier to learn from vectors with a skewed distribution on the entries. Therefore, we introduce a data preprocessing step that uses lattice-reduction methods to produce samples with smaller coefficients. In contrast to \salsa{}, \system{} starts with a linear number of samples, $m=4n$, and uses a novel subsampling procedure to generate many more LWE matrices for model training, deduplicated by the aforementioned preprocessing step. These design choices produce numerous non-duplicate samples with skewed entries, and we show that transformers can learn from such data better than from a large set of LWE samples without preprocessing. We also show that, compared to using data preprocessed on independent LWE pairs, the model learns equally well or better using preprocessed data on matrices subsampled from a linear number of LWE pairs.

Specifically, this work makes the following contributions: 
\begin{packed_itemize}
    \item {\bf Linear number of samples}: our method only requires a linear number of samples, $m=4n$ in practice.
    \item {\bf Data preprocessing:} we preprocess data with classical lattice reduction techniques (e.g. LLL~\cite{LLL} or BKZ~\cite{CN11_BKZ}), using small block size. 
    This helps transformers learn.
    \item {\bf Novel secret recovery:} we recover secret bits from the trained transformer, using its {\it cross-attention} mechanism. 
\end{packed_itemize}

We also improve the data encoding method, introduce rounding to reduce the size of the vocabulary the transformer must learn, improve the distinguisher secret recovery method, and introduce novel combined secret recovery methods. Finally, we compare \system{}'s performance to classical LWE attacks.

\para{Problem complexity.} \rev{Instances of cryptographic problems like LWE fall broadly into three buckets:
\begin{packed_itemize}
\item easy (solvable via exhaustive search);
\item medium-to-hard (requiring significant/unrealistic resources even for best known attacks);
\item standardized (believed secure). 
\end{packed_itemize}}

\rev{
\system{} considers {\bf medium-to-hard} problems and outperforms some lattice reduction attacks such as uSVP. For example, \system{} recovers secrets when $n=256$ and $h=31$.  In this setting, there are $2^{133}$ possible binary secrets, so brute force attacks are impossible. \system{} succeeds in $70$ hours with simple parallelization, while uSVP attacks run on the same machines succeed in $230-240$ hours (see Table \ref{tab:concrete_time}). In dimension $n=350$, $h=60$ ($2^{227}$ binary secrets), 
\system{} recovers sparse binary secrets in $\approx 250$ hours, whereas the uSVP attacks did not recover secrets. 
}

Overall, \system{} demonstrates a significant improvement over \salsa{}, further validating the possibility of ML-based attacks on LWE with sparse binary secrets. \system{} cannot (yet) break LWE schemes standardized by NIST,
which use larger dimension, smaller moduli $q$, and more general secret distributions.
But it has the potential to scale to these. Further research should explore and expand this line of work.

%% file: back.tex
\section{Background and Related Work}
\label{sec:back}

Before presenting \system{}, we first provide an overview of lattice cryptography and LWE, existing attacks on LWE, and information relevant to the machine learning techniques used in our attack.

\vspace{-0.2cm}
\subsection{Lattice-based cryptography}
\label{subsec:lwe}

Lattice-based cryptography is a major field in post-quantum cryptography. Three out of the four schemes selected by NIST~\cite{nist2022finalists} are lattice-based, and two  are based on a variant of LWE~\cite{Reg05}.

\para{Lattices.} An $n$-dimensional integer lattice is the set of all integer linear combinations of $n$ linearly independent vectors in~$\mathbb{Z}^n$. More formally, given $n$ vectors $\textbf{v}_1, \dots, \textbf{v}_n \in \mathbb{Z}^n,$ a lattice is the integer span $\Lambda = \Lambda(\textbf{v}_1, .. \textbf{v}_n) := \{ \sum_{i=1}^{n}{a_i \textbf{v}_i} \mid a_i \in \mathbb{Z}\}.$ The vectors $\textbf{v}_1, \dots, \textbf{v}_n$ are called a {\it basis} for the lattice $\Lambda$. The lattice $\Lambda$ inherits a norm simply by restriction of the Euclidean norm from $\mathbb R^n$ to $\Lambda$: any vector \rev{$\textbf{v} \in \Lambda$ has norm~$\| \textbf{v} \| = \sqrt{\textbf{v} \cdot \textbf{v}}.$} 

\para{Hard Lattice Problems.} Lattices give rise to several {\it hard} problems\textemdash problems for which the best known algorithms require exponential time in the dimension $n$ for both classical and quantum computers. The most famous and widely-studied is the {\it Shortest Vector Problem} (SVP): for a lattice $\Lambda$, find a nonzero vector $\textbf{v} \in \Lambda$ with minimal norm. Currently, the best algorithms for SVP take exponential space and time in~$n$~\cite{MV_SVP_exp}. This makes lattices attractive building blocks for post-quantum cryptography.

\para{Learning with Errors (LWE).} Many lattice-based cryptographic schemes leverage the ``Learning with Errors" problem, which is defined as follows. Fix a lattice dimension $n$, modulus~$q$, number of samples~$m$ and a narrow Gaussian probability distribution~$\chi$. The ``Learning with Errors'' (LWE) problem is to recover a secret vector ${\bf s} \in \mathbb{Z}_q^n$ given a collection of $m$ noisy samples  $({\bf a_i}, b_i)$, where ${\bf a_1, \dots, a_m } \leftarrow_R \mathbb Z_q^n$ 
are random vectors, and $b_i = {\bf a_i\cdot s} + e_i \bmod q$ 
are noisy inner products. The $e_i \in \mathbb Z_q$ are sampled independently from the error distribution $\chi$. A {\it LWE instance} is given by a matrix 
$(\bf A, \bf b) \it \in \ZZ_q^{m\times n}\times \ZZ_q^m$, where 
$\bf A$ is uniformly random in $\ZZ_q^{m \times n}$ 
and  ${\bf b} = {\bf A}\cdot {\bf s} + {\bf e} \bmod q$ is a column vector. The vector ${\bf s} \in \ZZ_q^n$ is the secret vector, and 
$ {\bf e}  \in \ZZ_q^m$ is an error vector with entries sampled from the probability distribution $\chi$. 
We call any of the pairs $(\textbf{a}_i, b_i)$, or equivalently any row of the matrix $({\bf A, b })$, an LWE {\it sample}.

\para{Hardness of LWE.}
In 2005, Regev demonstrated a worst-case quantum reduction from the SVP to LWE~\cite{Reg05}. 
Regev also showed that LWE-based cryptographic schemes were far more efficient than existing lattice cryptography methods. Later work demonstrated that LWE is classically as hard as worst-case SVP-like problems \cite{Pei09a,LM09_hardness_lwe_exp,BLPRS13_poly_modulus_hardness}. Hence, LWE is considered a solid foundation for (post-quantum) lattice cryptography.

\para{Real-world LWE-based cryptographic schemes.} 
LWE-based schemes are not only standardized for Post-Quantum Cryptography \cite{crys_kyber, crys_dilithium} and Homomorphic Encryption~\cite{HES}, but also allow for a range of cryptographic constructions beyond key exchange and signatures, including group signatures, secret sharing, and multi-party computation.
The NIST standardization competition received 23 entries proposing schemes based on lattice assumptions such as LWE.
In CRYSTALS-Kyber~\cite{crys_kyber}, the dimension is $n = k \times 256$ for $k = {2, 3, 4}$, \rev{where $k$ is a parameter of Kyber's module-LWE scheme}. The LWE-based signature scheme
CRYSTALS-Dilithium~\cite{crys_dilithium} uses similar size of~$n$. Both use secret vectors with small integer coordinates, centered around $0$. 
Another LWE-based NIST submission, LIZARD,
suggests LWE dimensions~$n$ from $544$ to $736$
\cite[Table 2]{Lizard}.

Homomorphic encryption schemes in publicly available libraries such as SEAL use dimension $n=512$ only for small computations, and generally require dimensions $n=1024, 2048$ and other powers of $2$ up to $2^{15}$. HE implementations commonly use binary or ternary secrets for efficiency (see~\cite{HES}), 
and many implementations propose using a sparse (binary) secret with Hamming weight $h << n$.  For instance, HEAAN uses $n = 2^{15},\,q = 2^{628}$, ternary secret and Hamming weight $64$~\cite{Cheon_hybrid_dual}. For more on the use of sparse binary secrets in LWE, see \cite{Albrecht2017_sparse_binary,Rachel_Player_sparse, heaan}. We focus on the case of a binary secret with Hamming weight $h$ and error distribution $\chi$, a centered Gaussian with $\sigma = 3$. $\sigma = 3.2$ is the typical choice for homomorphic encryption \cite{HES, SEAL, heaan, Rachel_Player_sparse}.

\vspace{-0.1cm}
\subsection{Attacks on LWE}
\label{sec:lwe_attacks}
The LWE problem is assumed to be exponentially hard to solve with classical~\cite{Pei09a,LM09_hardness_lwe_exp,BLPRS13_poly_modulus_hardness} or quantum~\cite{regev:quantum} algorithms.
Due to LWE's prominence as a hard problem in post-quantum cryptography,
a significant body of work has been devoted to attacking it. 

\para{Classical Attacks.} Most existing classical attacks on LWE leverage {\it lattice reduction} techniques, which reduce the problem to recovering the shortest vector in a lattice.
The LLL~\cite{LLL} algorithm runs in polynomial time in the dimension of the lattice (the optimized \fplll~\cite{fplll} implementation runs in time $O(n^4 \log(q)^2)$), but it recovers an exponentially bad approximation to the shortest vector.
LLL can be improved using the {\it block Korkine-Zolotarev} method (BKZ) by Schnorr~\cite{BKZ} and Schnorr-Euchner~\cite{schnorr_euchner}. \rev{The BKZ algorithm finds shortest vectors in projected lattices of dimension $k < n$, where  $k$ is referred to as the {\em block size}.} The BKZ approach relies on an exponential time sub-algorithm applied for increasing block sizes,
but can recover shorter vectors than LLL. 
The main 3 attacks used to estimate secure parameters for lattice-based cryptography are: the
uSVP, dual, decoding attacks, all of which require finding a short vector in a particular lattice arising from the LWE instance. 
The uSVP attack uses Kannan's embedding \cite{Kan87} to embed the problem into a lattice such that the (unique) shortest vector reveals the secret $s$. For concrete choices for this embedding, see \cite{CCLS}. 
The Homomorphic Encryption Standard~\cite[Section 2.1.2]{HES} describes the uSVP, dual, and decoding attacks in detail.

\para{\salsa{}: a machine learning based attack.} \salsa{}~\cite{wengersalsa} demonstrated the possibility of training machine learning (ML) models to attack LWE for sparse binary secrets. \salsa{} trained universal transformers to predict $b$ from input $\bf a$ and developed secret recovery techniques to extract the secret that is implicitly learned by the model. \salsa{} successfully recovered secrets for LWE problems with dimension $n \le 128$ and Hamming weight $\le 4$.

\subsection{Machine learning preliminaries}
\label{sec:ml_back}

 Here we provide a brief background and intuition behind the ML techniques used in our attack, \system{}. 

\para{ML basics.} The generic goal of machine learning is to compute a model $\mathcal{M}$ that maps an input $x \in \mathcal{X}$ to an output $y \in \mathcal{Y}$. The model $\mathcal{M}$  is computed via a {\it supervised training process}, during which it is shown samples $(x', y')$ such that $ x' \in \mathcal{X}' \subset \mathcal{X}$ and $ y' \in \mathcal{Y}' \subset \mathcal{Y}$. During this training process, the parameters $\theta$ of $\mathcal{M}$ are iteratively updated to minimize a predefined {\it loss function}, $l(\mathcal{M}, x',y', \Tilde{y})$, where $\Tilde{y}$ is the model's predicted output given input~$x'$ (e.g. $\mathcal{M}(x') = \Tilde{y}$), and $y'$ is the ground truth output. 

In \system{}, a model $\mathcal{M}$ is trained using LWE samples ($\bf a$, $b$) as~($x, y$) pairs. Hence, models are given an input vector $\bf a$ and asked to predict the LWE output $b$, where $b = {\bf a \cdot s} + e$. We use the cross-entropy loss function, a common choice in ML model training. For our model, we use the well-known {\em transformer} architecture. 

\para{Transformers.} Transformers were introduced in \cite{transformer17} for natural language processing (NLP) and machine translation. In recent years, they have been applied to a wide range of problems, from text and image generation \cite{radford2018improving,radford2019language, dalle2021} to image processing \cite{carion2020endtoend} and speech recognition \cite{DongSpeechTransformer}, where they now achieve state-of-the-art performance \cite{dosovitskiy2021image,Wang_2020}.
Transformers have also been proposed for problems in mathematics, like symbolic integration \cite{lample2019deep}, theorem proving \cite{polu2020generative}, and numerical computations \cite{charton2021linear}.
Transformers process sequences of tokens (in NLP, sequences of words, making up sentences). They combine a multi-head attention mechanism \cite{bahdanau2014} that takes care of relations between different tokens in the sequence, essentially ``decorrelating'' it, and a fully-connected neural network (FCNN), which processes the decorrelated sequences. More details about our transformers are in \S\ref{subsec:transformers}.

%% file: method.tex
\begin{figure*}[h]
    \centering
    \includegraphics[width=0.98\textwidth]{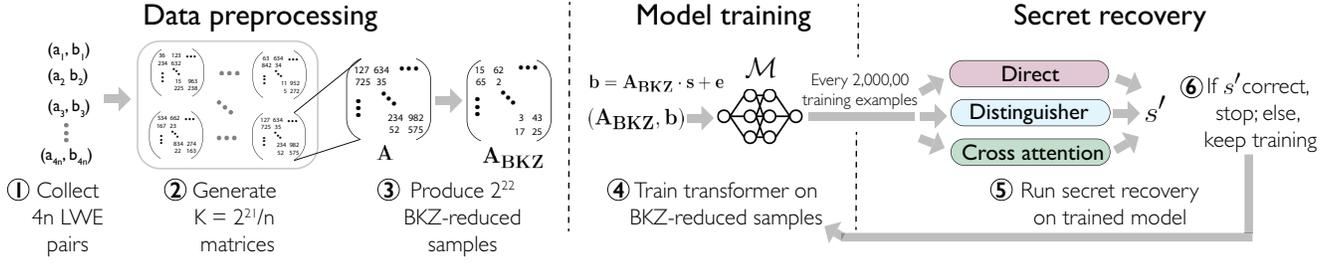}
    \vspace{-0.2cm}
    \caption{\textbf{An end-to-end overview of \salsys{}'s attack methodology.}
    \vspace{-0.1cm}
    }
    \label{fig:method_overview}
\end{figure*}

\vspace{-0.1cm}
\section{Introducing \salsys{}}
\label{sec:method}

Before diving into the details of \system{}'s methodology, we first present a high level overview of the attack. \salsys{} builds upon \salsa{} and progresses in three stages: (1) data preprocessing, (2) model training, and (3) secret recovery (see Figure~\ref{fig:method_overview}). 

Each run of \salsa{} \system{} targets LWE for a fixed dimension $n$, modulus $q$, binary secret $s$ with Hamming weight $h$, and error distribution $\chi$ with $\sigma = 3$. \salsa{} \system{} requires $m$ \textit{original LWE pairs}, sharing the same secret $s$. These are of the form $({\bf a_i}, b_i )$, with~$b_i={\bf a_i} \cdot {\bf s} + e_i$.
In real world situations, these samples must be collected. In experimental settings we choose $m= 4n$ and generate these samples randomly. After these parameters are fixed, the attack proceeds via the following three stages:

\para{(1) Data preprocessing.}
During this step, $n$ LWE pairs are randomly selected from the set of original samples and stacked into an $n \times n$ matrix $\textbf A$ and vector $\textbf b$ of length $n$. The matrix $\textbf A$ is processed using a basis-reduction algorithm (BKZ), and the same linear operations are performed on $\textbf b$. This creates
reduced LWE pairs with smaller norms but larger errors. This step is repeated to produce~$2^{22}$ reduced LWE pairs.

\looseness=-1 \para{(2) Model Training.} The reduced LWE samples $(\textbf a,b)$ are encoded as sequences of numbers, represented in base $B$, and used to train a transformer model $\T$. The model $\T$ learns to predict $b$ from $\textbf a$. 
Model training proceeds in \textit{epochs}, each using 2 million samples. The~4~million training data are shuffled randomly every 2 epochs. 

\para{(3) Secret Recovery.} At the end of each epoch, \salsys{} attempts to recover the secret using 3 techniques: direct, distinguisher, and cross-attention. The methods are used separately and can be combined to provide more secret guesses. Secret guesses are evaluated. Model training stops if the secret is recovered; else, another epoch begins.

%% file: attack.tex
\vspace{-0.1cm}
\section{Attack Methodology} 
\label{sec:attack}

Now, we provide a detailed description of \system{}, which progresses in the three stages outlined above.

\vspace{-0.2cm}
\subsection{Stage 1: data preprocessing}
\label{subsec:data_method}

\para{(1.1) \hspace{0.1cm} Collect LWE samples.} %
The \system{} attack begins by collecting a set of LWE samples with fixed parameters, as described in~\S\ref{sec:method}. \salsa{} assumed the attacker had access to $4,000,000 \approx 2^{22}$  LWE pairs ($\bf A, b$) with the same secret, since transformers, the model architecture used in both \salsa{} and \salsys{}, typically train on millions of examples. However, access to this many samples is unrealistic in practice. To mitigate this, \system{} introduces \tinyLWE{}, a technique that only requires~$m=4n$ LWE pairs -- linear in the dimension~$n$. Thus, the attack collects the $4n$ pairs and then runs \tinyLWE{}.

\para{(1.2) \hspace{0.1cm} Recombine to expand LWE sample set.}  The goal of \tinyLWE{} is to produce the large set of $4$ million samples required to train our models, from a small initial set of $m=4n$ LWE pairs.
Prior work~\cite{wengersalsa,LWEestimator} observed that a set of $m$ LWE pairs $(\textbf a_1,b_1) \dots (\textbf a_m,b_m)$ can always be expanded by considering the linear combinations $(\textbf a,b) = (\sum_i c_i \textbf a_i, \, \sum_i c_i b_i)$, with $c_i \in \mathbb{Z}$ and $\sum_i |c_i|$ small. We could, therefore, generate a large set of samples from a small initial set of LWE pairs by creating many such linear combinations. 

Unfortunately, LWE error is amplified by linear combinations: $e'=b-\textbf a \cdot \textbf s = \sum_i c_i e_i$, with $e_i$ the error in the original LWE sample. Assuming that the $c_i$ are centered, the standard deviation of error grows as the square root of the number of terms in the combination ($\sqrt m$ in the general case) times the standard deviation of the distribution of $c_i$ (which is~$\sqrt {C/3}$ if we assume the $c_i$ are uniformly distributed in~$[-C, C]$). 
\rev{In addition, the initial LWE error is further amplified by the data reduction step. So although the transformers used in \system{} can handle noisy data, generating samples via linear combinations would bring error to a level where training and secret recovery becomes very difficult.}

Instead of linear combinations, \system{} uses {\it subsampling}. Subsets of $n$ out of the $m$ original LWE samples, $(\textbf a_{j_1},b_{j_1}), \dots (\textbf a_{j_n},b_{j_n})$, are randomly selected, and arranged in a matrix~$\textbf A$, with rows $\textbf a_{j_1}$ to~$\textbf a_{j_n}$.  
Because the original LWE pairs are merely copied into~$\textbf A$, the associated noisy inner products have the same error distribution as the original samples. This technique produces up to ${4n \choose n}$ unique matrices ($\approx {9.48}^n \cdot 0.46/\sqrt n)$). In \system{}, we use subsampling to generate about $2^{21}/n$ matrices, which, after the reduction step described next, results in about $2^{22}$ reduced LWE pairs. Subsampled matrices often have rows in common, but we experimentally observe that, after reduction, there are almost no duplicate vectors. For $n=80$, we counted one duplicate in $50,000$ examples; for $n \geq 150$, we found no duplicates in $4$ million examples.

\rev{Note that \emph{subsampling} is different from \emph{batching}. Subsampling allows us to generate a training set of $4$ millions examples from only $4n$ LWE samples. This is accomplished in the preprocessing step, which performs data reduction on subsets of the $4n$ original samples. Batching, on the other hand, takes place during training, when computing the gradients of the loss function, that are then used to optimize the model. Instead of computing gradients on a single training example, batching averages them over many examples, allowing for faster training and better estimate of gradients. \system{} uses batches of $128$ examples.
}

\para{(1.3) \hspace{0.1cm} Reduce samples.} After sampling and recombination, the attack runs a final {\em reduction} step to make the LWE samples more amenable to model training. The motivation for this step comes from experimental observations in \salsa{}. \salsa{} could only recover binary secrets with low Hamming weights: up to $4$ non-zero bits in the secret. However, the \salsa{} authors observed~\cite[Table 4]{wengersalsa} that, if the coordinates of the samples $\bf a$ used to train the model were bounded by $\alpha q$, with $\alpha < 0.6$, binary secrets with Hamming weights up to $15$ could be fully or partially recovered for $n=50$.  We confirmed this result for larger dimensions, and different restrictions on $\textbf a$ (e.g.~$a_i \le \alpha q$ for different $\alpha$). Unfortunately, in practical settings, the coordinates of $\textbf a$ are sampled from a uniform distribution over~$\mathbb{Z}_q$, making this technique useless for real world attacks. 

\system{} turns this observation into a practical attack technique by leveraging existing {\em lattice reduction} methods. Such methods reduce the size of the coordinates of LWE samples naturally, yielding the same effect as the \salsa{} $\bf a$-limiting technique. We find experimentally that reducing LWE samples via these methods before model training allows recovery of secrets with much higher Hamming weights. The reduction technique proceeds as follows. 

Given $n$ LWE samples, stored as the rows of a $n \times n$ matrix $\textbf A$, and a corresponding vector $\textbf b$ of noisy inner products with a fixed secret $\textbf s$, we can create a matrix $\textbf A'$ with smaller entries than $\textbf A$ by applying standard basis-reduction algorithms like LLL \cite{LLL} and BKZ \cite{BKZ} to $\Lambda$, the $n$-dimensional lattice defined by the rows of $\bf A$. In \system{}, we run BKZ (from the \fplll{} package \cite{fplll}) 
on the matrix:
\[\begin{bmatrix}
\omega\cdot\mathbf{I}_n & \mathbf{A}_{n\times n} \\
0 & q\cdot \mathbf{I}_n
\end{bmatrix}, \]
with $\omega \in \mathbb{Z}$ an error penalization parameter, discussed below. Since the BKZ reduction is a change of basis, it is a linear transformation, which we can represent as 
$\begin{bmatrix} \mathbf{R}_{2n\times n} & \mathbf{C}_{2n\times n} \end{bmatrix}.$ The BKZ reduction
can be written as a matrix multiplication
\[\begin{bmatrix} \mathbf{R}_{2n\times n} & \mathbf{C}_{2n\times n} \end{bmatrix} 
    \begin{bmatrix}
        \omega \cdot\mathbf{I}_n & \mathbf{A}_{n\times n} \\
        \mathbf{0} & q\cdot \mathbf{I}_n
    \end{bmatrix}
= \begin{bmatrix} \omega \cdot\mathbf{R} & \mathbf{RA}+q\mathbf{C} \end{bmatrix},\]
with matrices
$\mathbf{R}$ and $\mathbf{C}$ chosen so that $\begin{bmatrix} \omega \cdot\mathbf{R} & \mathbf{RA}+q\mathbf{C} \end{bmatrix}$ has $2n$ rows with small norms. The matrix $q\mathbf{C}$ adds an integer multiple of $q$ to each entry in $\mathbf{RA}$, so that all entries are in the range $(-q/2, q/2)$.

Applying the linear transformation $\textbf R$ to $\textbf b$, we create a new LWE instance $(\bf RA, \bf Rb)$ 
with the same secret $\bf s$ and smaller coordinates $\bf RA$ but {\it a different error distribution}.
Let $\textbf e = \textbf b -\textbf A \cdot \textbf s$ be the initial LWE error. After reduction, the error becomes $\textbf e' =  \mathbf{Rb - (RA)\cdot s = R(b-A\cdot s) = Re}$. Thus, as $\textbf R$ entries grow, LWE error is amplified. All computations are performed $\bmod$ $q$.

Error amplification can be controlled by the error penalization parameter $\omega$. Recall that BKZ computes $\mathbf{R}$ and $\mathbf C$ so that the norms of the rows of $\begin{bmatrix} \omega \cdot\mathbf{R} & \mathbf{RA}+q\mathbf{C} \end{bmatrix}$ are small. A large $\omega$ encourages small entries in the rows of $\mathbf{R}$ but hinders the norm reduction of $\mathbf{RA}+q\mathbf{C}$, and therefore limits the reduction of $\bf A$ coordinates. The choice of $\omega$ controls a trade-off between the amount of reduction of $\textbf a$ we can achieve, and the amount of additional noise which gets injected in the transformed samples. In practice, we set $\omega=15$.

When more than $n$ pairs are available (e.g. the million of pairs produced by Step 1.2), they are divided into batches of $n$ and processed as above. Thus, $n$ LWE pairs are transformed into a matrix $\textbf{RA}$ with~$2n$ rows, which produces $\approx 2n$ reduced LWE samples (for $n\geq 256$, we observe about $1\%$ zero rows; this fraction is larger for smaller~$n$).

{\it Note on reduction algorithm choice.} \hspace{0.1cm} We experimented with two standard basis-reduction algorithms: LLL and BKZ. Note that our objective is not to find the shortest vector in the lattice defined by $\bf A$ (the traditional goal of LLL/BKZ), but to transform~$\bf A$ into a matrix with smaller coefficients. Experimentally, we find that BKZ with small block size ($\beta=16-20$) achieves better reduction than LLL. BKZ speed-ups, such as BKZ2.0 \cite{CN11_BKZ}, do not seem to result in improved reduction.
For BKZ, the block sizes needed to achieve reduction in
\system{} are significantly smaller than the block sizes that would be required to perform a lattice-reduction attack on problems of the same dimensions (see also \S~\ref{sec:compare}). 

\vspace{-0.15cm}
\subsection{Stage 2: model training}
\label{subsec:transformers}

After the data is prepared, the attack enters the model training stage. Although there are no sub-stages to model training, here we break down the model training step into several components: data encoding, model architecture choice, and the training itself.

\para{Encoding LWE pairs}. Prior to training, \system{} encodes the LWE samples (i.e. $({\bf a}, b)$ pairs) as sequences of tokens that the transformer can process. After encoding, the integer coordinates of $\textbf{a}$ and~$b$ are represented as two digit numbers in base $B$ (with~$B \geq \sqrt q$).
Our experiments with different values of $B$ (see \S~\ref{subsec:base_ablation}) suggest that large values of $B$, which limit the most significant digit of $a_i$ and~$b$ to a small number of values (i.e. $B\approx q/k$ with $k$ small), allow for better performance. In our experiments, we use $B = \lfloor q/k \rfloor$ with~$k = 2\cdot \ceil{\frac{n}{100}}+2$.

This creates a problem for large dimensions. The large values of $q$ and $B$ (for $n\geq20$ we have $q>100,000$ and $B>16,600$) result in large token vocabularies, which are difficult to learn for a transformer trained on $4$ million LWE pairs only. To mitigate this, we encode the lowest digits of $\textbf a$ and $b$ into $B/r$ \textit{buckets} of size $r$ (i.e. integer divide them by~$r$). The value $r$ is chosen so that the overall vocabulary size $B/r < 10,000$ (see Table~\ref{tab:params}).
The use of buckets helps train models for large $n$ but it also causes a loss of precision in the values of $\textbf a$ and $b$. We believe the impact on performance is limited, because the low bits of~$\textbf{a}$ and~$b$ that are rounded off by buckets are those most corrupted by LWE error.

\para{Model architecture.} As noted previously, \system{} uses a transformer model architecture. This architecture, summarized in Figure~\ref{fig:archi}, is strongly inspired by SALSA \cite{wengersalsa}. Following \cite{transformer17}, it uses a sequence-to-sequence (seq2seq) model \cite{seq2seq2014}, composed of two transformer  stacks -- an encoder and a decoder -- connected by a cross-attention mechanism. The encoder processes the input sequence, the coordinates of $\textbf{a}$, represented as sequences of digits. The discrete input tokens are first projected over a high-dimensional space (we use dimension $d=1024$) by a Linear Embedding Layer with trainable weights (i.e. embedding is learned during training). The resulting sequence is then processed by a single-layer transformer: a self-attention layer with $4$ attention heads, and a FCNN with one hidden layer of $4096$ neurons.

The decoder is an auto-regressive model. It predicts the next token in the output sequence, given already decoded output and the input sequence processed by the encoder. Initially, the decoder is given a beginning of sequence token (\texttt{BOS}), and predicts $b^*_1$, the first digit of $b$. It is then fed the sequence \texttt{BOS}, $b^*_1$, and decoding proceeds until the end-of-sequence token (\texttt{EOS}) is output.

Decoder input tokens are encoded as $512$-dimensional vectors via a trainable embedding (which also decodes transformer output). The decoder has two layers. First, a shared layer (as in \cite{dehghani2018universal}), which is iterated through $8$ times, feeds layer output back into its input. This recurrent process is controlled by a copy-gate mechanism \cite{csordas2021neural}, which decides whether a specific token should be processed by the shared layer or just copied \textit{as is}, skipping the next iteration.
After $8$ iterations, the output of the shared layer is fed into a ``regular'' transformer layer. Finally, a linear layer processes the decoder output and computes the probabilities that any word in the vocabulary is the next token. The largest probability is selected via a softmax function (a differentiable counterpart of the max function).

Decoder layers are connected to the encoder via a {\em cross-attention} mechanism with $4$ attention heads. In each head, the output of the encoder $E=(E_i)_{i\in \mathbb{N}_l}$ (with $l$ the input sequence length) is multiplied by two trainable matrices, $W_K$ and $W_V$, yielding the \textit{Keys}~$K=W_KE$ and \textit{Values} $V=W_VE$. The 512-dimensional vector to be decoded,~$D$, is  multiplied by a matrix $W_Q$, yielding the \textit{Query} $Q=W_Q D$. The~$l$~scores are calculated from the query and keys:
\[\text{scores}(E,D)= \text{Softmax}((W_Q D)(W_K E)^T).\] 
The scores measure how important each encoder input element is when decoding $D$
(i.e. computing $b$). 
The cross-attention value for this head is the dot product of scores and values. The values of different heads are then 
processed by a final linear layer.
Cross-attention scores quantify the relation between input positions and output values. \system{} uses them to recover the secret bit by bit.

\input{figs/transformer_pic.tex}

\para{Model training.} After encoding the samples, the attacker trains the transformer $\mathcal{M}$ to predict $b$ from $\bf a$. \system{} frames this as a supervised multi-classification problem, i.e. minimizing the loss:
\begin{align}\label{eq:problem}
    \hspace{-.3cm}\min_{\theta\in\Theta} \sum_{i=1}^{N} \sum_{j=1}^{K} \sum_{k=1}^{V} \mathbf{1}[y_{i}[j]=k-1]\frac{e^{\mathcal{M}(x_i)[j,k]}}{\sum_{k'=1}^{V} e^{\mathcal{M}(x_i)[j,k']}},
\end{align}
where $\mathcal{M}(x_i)\in\mathbb{R}^{K\times V}$ are model logits evaluated at $x_i$, $\theta\in\Theta$ are the model parameters, $N$ the training sample size, $K=2$ the output sequence length and $V=B/r$ the vocabulary size.

Solving \eqref{eq:problem} requires minimizing the cross entropy between model predictions $\T(\bf a)$ and the ground truth $b$, over all tokens in the output sequence. Alternatively, one could define this as a regression problem, but we believe classification is better adapted to the modular case. Prior works confirm that reformulating regression as classification leads to state-of-the-art performance \cite{rothe2015dex,rogez2017lcr,akkaya2019solving,schrittwieser2020mastering}. 

Training proceeds via batches of $n_b=128$ examples. The cross-entropy loss $\mathcal L(\T,\textbf a,b) $ is computed over all batch examples, and gradients $\nabla\mathcal L$ 
 are calculated with respect to the model parameters (via 
\textit{back-propagation}). Model parameters are then updated using the Adam optimizer \cite{kingma2014adam}, by $\text{lr} \nabla  \mathcal L$. The learning rate, $\text{lr}$ is set to $10^{-5}$, except during the $1000$ first optimizer steps, where it is increased linearly from $10^{-8}$ to $10^{-5}$.  Every $2$ million examples (an \textit{epoch}), model performance is evaluated on a held-out sample, and \system{} attempts to recover the secret.  If it fails, another epoch begins.

\subsection{Stage 3: secret recovery}
\label{subsec:recovery}

After every training epoch, \system{} attempts secret recovery. The intuition behind secret recovery is that if a model $\mathcal{M}$ can predict $b$ from $\textbf{a}$ with higher-than-chance accuracy, then $\mathcal{M}$ must somehow ``know'' the secret key $\textbf s$, and we can recover $\textbf s$ from $\mathcal{M}$. \salsys{} uses $3$ methods\textemdash \textit{cross attention}, \textit{direct recovery}, and \textit{distinguisher} \textemdash for recovery. These can be combined for greater accuracy.

\rev{In this section, we assume that the attacker knows the Hamming weight $h$ of the secret to be recovered. This is the only part of the attack where this  assumption is made. If $h$ is not known, then secret recovery is run for increasing values of $h$ until the secret is found.}

\para{Cross-Attention.} In this novel recovery method, \system{} guesses the secret from the parameters of $\mathcal{M}$ by leveraging the cross-attention scores of the first decoder layer (see Figure~\ref{fig:archi} and \S~\ref{subsec:transformers}). Intuitively, the cross-attention score measures the relevance of input tokens (i.e. coordinates of $\textbf{a}$) for the computation of $b$. 
Since $b=\textbf{a}\cdot\textbf{s}+e$, the coordinates of $\textbf{a}$ that correspond to the $0$ bits of~$\textbf{s}$ have no impact on $b$. On the other hand, the coordinates associated to the $1$ bits in $\textbf s$ have an impact proportional to their value. Therefore, high cross-attention scores should be found for the input positions that correspond to $1$s in the secret.

To run this method, \system{} evaluates the trained transformer on a test set of $10,000$ reduced LWE samples and sums the cross-attention scores of all heads. Since $\textbf a$ has $n$ coordinates encoded on $2$ tokens, this produces a $2n$-dimensional vector, from which the odd positions are kept (i.e. the high digits of $\bf a$ coordinates), generating an $n$-dimensional score vector $V$. A secret guess  
$\textbf s'$  is then produced by setting the $h$ largest coordinates of $V$ to one, and the rest to $0$.

\para{Direct recovery.} \system{} uses the same direct recovery method as \salsa{}. This technique leverages trained transformers' ability to generalize on inputs not seen during training. The trained model is evaluated on special vectors~$\textbf{a}$ with one non-zero coordinate: $\textbf a=K \textbf e_i$ with $\textbf e_i$ the i-th standard basis vector 
and $K \in \mathbb{Z}_q$. For these vectors, since $b=\textbf a \cdot \textbf s + e$, and $e$ is small, $b \approx 0$ if the i-th bit in the secret $s_i=0$, and $b \approx K$ if~$s_i=1$ (see~\cite{wengersalsa} for details). In practice, different $K_j$ are chosen, and the transformer is run on $K_j \cdot \textbf{e}_i$ for $i= 1, \dots, n$, identifying potential $1$-bits in the secret as above and producing a secret guess for each $K_j$.

To obtain a score for each bit to be used in combination methods, for each index $i$, we sum the resulting values of $\T(K_j \cdot \textbf{e}_i)$ (or, equivalently, take the mean). We then guess the secret $\textbf s'$ by assuming that the~$h$ largest coordinates are $1$ and the rest are $0$.

\para{Distinguisher.} \system{}'s version of distinguisher recovery improves upon that of \salsa{}. 
The general idea is that if the~$i$-th bit of the secret $s_i=0$ and ${\textbf e}_i$ is the $i$-th standard basis vector, then the model should predict close values for $\textbf a$ and~$\textbf a+K \cdot \textbf e_i$. \salsa{}'s distinguisher took a LWE sample $(\textbf a, b)$ and compared $b$ with the model prediction $b' = \T({\bf a} + K \cdot {\bf e}_i)$ for some random $K \in \mathbb{Z}_q$. This presupposed relatively high model accuracy, i.e.~$\T(\textbf a) \approx b$, which rarely happens in practice. In \system{}, $b'=\T({\bf a} + K \cdot {\textbf e}_i)$ is compared to $\T (\textbf a)$ instead of to $b$. The rest of the method is unchanged, other than implementation improvements. The secret is guessed by setting the $h$ highest-scoring secret bits to $1$, and the rest to $0$.

This improved method has two benefits. First it exploits trained model consistency without requiring prediction accuracy. In practice, this means recovery can happen earlier during training, when model prediction accuracy is low. Second, it does not need additional LWE samples $(\textbf a, b)$ (as was the case in \salsa{}), and can be run from randomly generated $\textbf a$. This reduces the number of LWE samples necessary for the attack.

This recovery method relies on a large number of model inferences, which can make it very slow for large dimension. To increase its speed, we use the same $\bf a$ across all secret bits $s_i$, halving the number of inferences relative to those required in \salsa{}.

\para{Combined secret recovery.} Each recovery method outputs a score for every bit in the secret. The secret guess is computed by setting the $h$ bits with the largest scores to $1$ (and the other bits to $0$). By combining the scores from different methods, we create four additional techniques, which can sometimes can recover secrets when individual methods fail. The combined methods are as follows:

\begin{itemize}[leftmargin=*, itemsep=1pt, topsep=0.5pt, parsep=1pt] 
\item {\em Aggregated rank.} The bit scores produced by each method are sorted from largest to smallest, and replaced by their rank. The $h$ bits with the highest ranks ({Highest Rank}) or highest summed ranks ({Sum Rank}) are set to $1$.

\item {\em Aggregated normalized scores.} The bit scores produced by each method are normalized to $[0,1]$. The $h$ bits with the maximum normalized scores ({Max Normalized}) or the highest sum of normalized scores ({Sum Normalized}) are set to $1$. 
\end{itemize}

\noindent These combination rules essentially amount to setting secret bits to $1$ for bit positions where all, some, or any of the secret recovery methods have a high score.
We use aggregated scores from all subsets of the secret recovery methods. Other combination rules could be considered. These mixing techniques are cheap to implement, because they do not require additional model inferences.

\para{Checking correctness.}
Recovery methods make guesses $\textbf s'$ about the (unknown) secret $\textbf s$. The test for whether $\textbf s'=\textbf s$ was introduced in \salsa{}. On a test sample of $N_{\text{test}}$ LWE pairs $(\textbf a_i, b_i)_{1 \leq i \leq {N_{\text{test}}}}$, compute $b_i' = \textbf a_i \cdot \textbf s' $, and consider the distribution $r$ of $r_i = b_i' - b_i \mod q$. If $\textbf s' = \textbf s$, then $r \approx e$, the LWE error, with standard deviation~$\sigma$. If $\textbf s' \neq \textbf s,$ then $r$ will be approximately uniformly distributed over $\mathbb{Z}_q$, with standard deviation~$\sigma' = q/\sqrt{12}$. By estimating~$\sigma'$ on a large set of samples, one can verify $\textbf s'=\textbf s$ to any confidence level. 

This test can be performed on the original set of LWE samples collected by the attacker, e.g. with $N_\text{test}=m = 4n$. In \S\ref{app:verif} we statistically analyze this verification technique and demonstrate that this sample size is sufficient for all lattice dimensions $n \geq 80$.

%% file: figs/transformer_pic.tex
\definecolor{DarkGray}{HTML}{68904d}
\definecolor{Orange}{HTML}{ffae42}
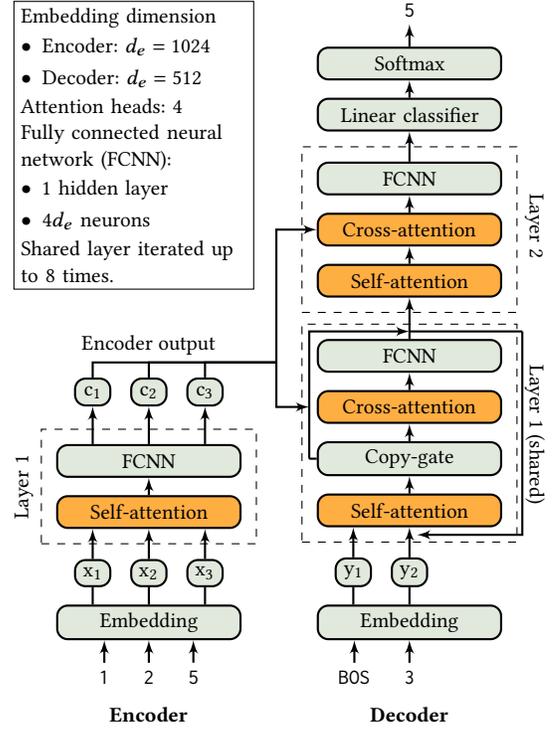
\begin{figure}[h!]
\small
\centering
\begin{tikzpicture}[
    box/.style={rectangle,draw,fill=DarkGray!20,node distance=1cm,text width=8em,text centered,rounded corners,minimum height=1.5em,thick},
    box2/.style={rectangle,draw,fill=DarkGray!20,node distance=1cm,text width=1em,text centered,rounded corners,minimum height=0.5em,thick},
    arrow/.style={draw,-latex',thick},
  ]
  \node [box] (ffne) {FCNN};

  \node [box2,above=0.5 of ffne] (tok2) {c$_2$};
  \node [box2,left=0.25 of tok2] (tok1) {c$_1$};
  \node [box2,right=0.25 of tok2] (tok3) {c$_3$};
  \node [above=0.2 of tok2]{Encoder output};
  \node [box,below=0.23 of ffne,fill=Orange] (selfatt){Self-attention};  
  \node [label={[rotate=90]160:Layer 1},rectangle,draw,dashed,inner sep=0.7em,fit=(selfatt) (ffne)] (encoder1) {};

  \node [box2,below=0.4 of selfatt] (se){x$_2$};
  \node [box2,left=0.25 of se] (se2){x$_1$};
  \node [box2,right=0.25 of se] (se3){x$_3$};
  \node [box,below=0.23 of se] (embed) {Embedding};
  \node [below=0.3 of embed] (inp2) {\texttt{2}};
\node [right=0.25 of inp2] (inp3) {\texttt{5}};
\node [left=0.25 of inp2] (inp1) {\texttt{1}};
\path[arrow] (inp2) -- (embed);
\path[arrow] (inp3) -- (embed.south -| inp3);
\path[arrow] (inp1) -- (embed.south -| inp1);
\node [below=0.1 of inp2]{\textbf{Encoder}};

\node [box,right=1.0 of embed](emb2){Embedding};

\node [below=0.3 of emb2](enc2){\texttt{3}};
\node [left=0.25 of enc2](enc3){\texttt{BOS}};
\node [below=0.1 of enc2]{\textbf{Decoder}};

\node [box2,above=0.23 of emb2](em22){y$_2$};
\path[draw,thick] (emb2) -- (em22);
\path[arrow] (enc2) -- (emb2);
\path[arrow] (enc3) -- (emb2.south -| enc3);

\node [box2,left=0.25 of em22](em21){y$_1$};
\path[draw,thick] (emb2.north -| em21) -- (em21);

\node [box,above=0.4 of em22,fill=Orange] (selfatt2){Self-attention};
\path[arrow] (em22) -- (selfatt2);
\path[arrow] (em21) -- (selfatt2.south -| em21);

\node [box,above=0.23 of selfatt2] (gate){Copy-gate};
\path[arrow] (selfatt2) -- (gate);

\node [box,above=0.23 of gate,fill=Orange] (XattD){Cross-attention};
\path[arrow] (gate) -- (XattD);

\node [box,above=0.23 of XattD] (ffn) {FCNN};
\path[arrow] (XattD) -- (ffn);

\node [label={[rotate=270]35:Layer 1 (shared)},rectangle,draw,dashed,inner sep=0.7em,fit=(selfatt2) (ffn)] (decoder2) {};
\node [box,above=0.55 of ffn,fill=Orange] (selfatt3){Self-attention};
\path[arrow] (ffn) -- (selfatt3);

\node [box,above=0.23 of selfatt3,fill=Orange] (crossatt2){Cross-attention};
\path[arrow] (selfatt3) -- (crossatt2);

\node [box,above=0.23 of crossatt2] (ffn2) {FCNN};
\path[arrow] (crossatt2) -- (ffn2);
\node [label={[rotate=270]22:Layer 2},rectangle,draw,dashed,inner sep=0.7em,fit=(selfatt3) (ffn2)] (decoder2) {};
\node [box,above=0.4 of ffn2] (linear){Linear classifier};
\path[arrow] (ffn2) -- (linear);
\node [box,above=0.23 of linear] (softmax){Softmax};
\path[arrow] (linear) -- (softmax);
\node[above=0.3 of softmax](output){\texttt{5}};
\path[arrow] (softmax) -- (output);

\coordinate[above right =0.2 and 0.7 of tok3](conn);
\path[draw,thick] (tok1) |- (conn);
\path[draw,thick] (tok2) |- (conn);
\path[draw,thick] (tok3) |- (conn);

\coordinate[left=0.1 of gate](lg);
\path[draw,thick] (gate) -- (lg);
\coordinate[above=0.1 of ffn.north](mw2);
\path[arrow] (lg) |- (mw2);
\coordinate[right=1.5 of mw2](mw3);
\path[draw,thick] (mw2)-- (mw3);
\coordinate[below right =0.1 and 0.1 of selfatt2.south](mw);
\path[arrow] (mw3) |- (mw);
 
\path [arrow] (ffne.north -| tok1) -- (tok1);
\path [arrow] (ffne) -- (tok2);
\path [arrow] (ffne.north -| tok3) -- (tok3);
\path [arrow] (selfatt) -- (ffne);
\path [arrow] (se) -- (selfatt);
\path [arrow] (se2) -- (selfatt.south -| se2);
\path [arrow] (se3) -- (selfatt.south -| se3);
\path[draw,thick] (embed.north -| se2) -- (se2);
\path[draw,thick] (embed) -- (se);
\path[draw,thick] (embed.north -| se3) -- (se3);
\coordinate[left=0.1 of XattD.west](mw4);
\path[arrow] (conn) |- (mw4);
\path[arrow] (conn) |- (crossatt2.west);

\node[draw,text width=3cm] at (-0.2,4.2) {Embedding dimension\\
\begin{itemize}[leftmargin=*, itemsep=1pt, topsep=1pt, parsep=1pt]
\item Encoder: $d_e=1024$ \\
\item Decoder: $d_e=512$ 
\end{itemize}
Attention heads: $4$

Fully connected neural network (FCNN):\\ 
\begin{itemize}[leftmargin=*, itemsep=1pt, topsep=1pt, parsep=1pt]
\item 1 hidden layer \\
\item $4d_e$ neurons \\
\end{itemize}
Shared layer iterated up to $8$ times.
};
\end{tikzpicture}
\vspace{-0.1cm}
\caption{\small\textbf{Our transformer architecture.} }\label{fig:archi}
\end{figure}

%% file: results.tex
\vspace{-0.15cm}
\section{\salsa{} \system{}'s Performance}
\label{sec:results}

We now evaluate \system{}'s performance over a variety of parameter settings for lattice dimension, modulus size, Hamming weight, and number of samples. 
All \system{} experiments are based on the following choices, with the exact parameters used in our experiments listed in Table~\ref{tab:params}. Other details are in \S\ref{sec:attack}.

\subsection{Experimental settings}\label{subsec:exp_setting}
\begin{itemize}[leftmargin=*, itemsep=0.5pt, topsep=1pt, parsep=1pt]
    \item For each $n$, the modulus $q$ is selected after consulting Table 1 in \cite{CCLS}. We set our $q$ such that $\text{log}_2 q$ is smaller than the smallest successful lattice-reduction attack reported there (see Table~\ref{tab:params}). A smaller $q$ makes attacks more difficult.
    \item The error in the original LWE samples is sampled from a discrete Gaussian distribution, centered at $0$, and with $\sigma=3$, a common choice when LWE is used in homomorphic encryption~\cite{Albrecht2017_sparse_binary, HES}. 
    \item We consider binary secrets with sparsity $h/n \approx 10\%$ or larger. For each $n$ where we evaluate \system{}, we show results for seven different Hamming weights $h \approx n/10$ to confirm repeatability.
\item The attack starts with a set of $4n$ randomly generated samples ($\textbf a ,b$) with fixed $n$, $q$, sparse binary $\bf s$, and $\sigma$. 
\item For the BKZ reduction step (\system{} stage 1.3), we use $\omega=15$ as the error penalization parameter for all $n$. Block size and the LLL parameter $\delta$ in \fplll{} are set to $20$ and $0.99$ for all $n\leq 200$. To keep preprocessing times reasonable, we decrease these values for $n=256$, $300$ and $350$ (see Table~\ref{tab:norm_reduction}).
\item For each $n$ and $q$, we perform the preprocessing step on random matrices $A$ once and use that reduced data for experiments with different secrets. 
\end{itemize}

\begin{table}[h!]
\small
\centering
\begin{tabular}{ccccccc}
\toprule
$n$ &$q $ & $\log_2 q$& $\delta$ &$\beta$ & base & $r$ \\ \midrule
80  & 113        & 7  & 0.99 & 20   & 29        & 1      \\
150 & 6421       & 13 & 0.99 & 20   & 1071      & 1      \\
200 & 130769     & 17 & 0.99 & 20   & 21795     & $2^2$      \\
256 & 6139999    & 23 & 0.96 & 18   & 767500    & $2^7$    \\
300 & 94056013   & 27 & 0.96 & 16   & 11757002  & $2^{11}$   \\
350 & 3831165139 & 32 & 0.96 & 14   & 383116514 & $2^{16}$  \\ 
\bottomrule
\end{tabular}
\vspace{0.1cm}
\caption{\small\textbf{\system{} parameters.} $n$: dimension, $ q$: modulus, $\delta$: delta-LLL (BKZ), $\beta$: block-size (BKZ), base: encoding base, $r$: bucket size for encoding.
}
\vspace{-0.8cm}
\label{tab:params} 
\end{table}

\subsection{Overall Performance}
\label{subsec:best_results}

A summary of \system{}'s results is shown in Table~\ref{tab:success_epochs}, which records \system{}'s success for various dimensions $n$, modulus $q$, and Hamming weight $h$. We run multiple experiments for each parameter setting, and report the number of successes/attempts, as well as the model training epochs at which successful secret recoveries occurred. For example, in dimension $n=350$, we recovered a secret with $h=60$ in one out of five trials (each trial has a different secret).  In that case, the recovery happened in training epoch $38$.
 
\para{Effect of dimension $n$.} For dimensions up to $300$, \system{} consistently recovers LWE secrets with sparsity $h/n \approx 10\%$, a significant improvement over \salsa{}. For $n=350$, \system{} can recover secrets with Hamming weight $h=60$, sparsity $\approx 17\%$. 
We believe this improved performance is due to the preprocessing parameters used for $n=350$ (\S\ref{subsec:ablate_preproc}). This suggests that harder LWE problems, with dimension $n=350$ but smaller $q$, could be solved with this architecture for $h\approx 0.1n$.

For all dimensions, \system{} succeeds for smaller values of the modulus $q$ than those for which the concrete, classical lattice attacks in~\cite[Table 1]{CCLS} can recover secrets with BKZ blocksize roughly $40$. In our experiments, we use a fixed $q$ for each dimension, to avoid running the costly preprocessing step multiple times. Evaluating performance at varying $q$ for a fixed $n$ is important future work.

\para{Effect of Hamming weight $h$.} For each dimension $n$, we evaluate \system{} on secrets with a range of Hamming weights. For each~$n$, there is a ``cutoff'' Hamming weight, above which \system{} did not successfully recover the secret in these runs. This is expected, because increasing Hamming weight makes the problem more difficult. Table~\ref{tab:success_epochs} presents the cutoff value in bold, along with the number of successfully recovered secrets for each Hamming weight.

\input{tables/table_success.tex}

\begin{figure}[t]
\small
    \includegraphics[width=0.45\textwidth]{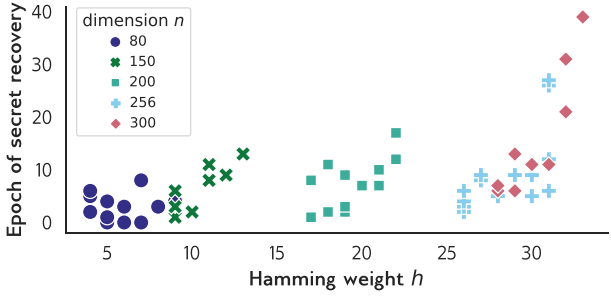}
    \vspace{-0.1cm}
    \caption{\small\textbf{Number of training epochs before secret recovery for $80 \le n \le 300$.} For different dimensions and Hamming weights. We omit $n=350$ results for space reasons.}
    \label{fig:tiny_lwe_epochs}
    \vspace{-0.2cm}
\end{figure}

\para{Required training duration.} Figure~\ref{fig:tiny_lwe_epochs} shows the number of epochs needed for secret recovery for $80 \le n \le 300$ and different values of~$h$. 
Whereas $66\%$ of successful secret recoveries occurred during the first $10$ epochs, the number of epochs before recovery increases with $n$ and $h$. For $n=80$, about $75\%$ of successful experiments succeed by epoch $4$. $8$ epochs are needed for $75\%$ of experiments to succeed for $n=150$, and $13$ epochs for $n=200,256, 300$.

For a given dimension, the number of epochs required for secret recovery varies a lot from one experiment to another. For dimension~$256$ and Hamming weight $31$, different secrets need between $6$ and~$27$ epochs. For dimension $350$, a secret with Hamming weight~$59$ is recovered after $18$ epochs, while secrets with $h=58$ and $60$ need~$39$ and~$38$.
 
We believe that, for a given secret $\textbf s$, certain distributions of the coordinates of $\textbf a$ help the transformer learn $\bf s$. The proportion of such points $\textbf a$ in the training sample varies for different secrets, making some harder to recover, and necessitating longer training.

Another explanation for the variations in training length is the random initialization of transformer parameters, discussed in  \S~\ref{subsec:seeds_ablate}. For a given secret, running several experiments, with different initializations, may reduce the number of epochs required for recovery. 

\para{Required LWE sample size.} \system{} relies on the \tinyLWE{} subsampling technique introduced in \S\ref{subsec:data_method} to recover secrets from only $4n$ initial LWE samples. By comparison, \salsa{} used $4$ million LWE samples. Table~\ref{tab:tiny_vs_lwe} compares the performance of \system{} (using \tinyLWE{}), with an equivalent attack using $2.2$ million collected LWE samples, which we call \lwe{}. For both sampling approaches \textemdash\tinyLWE{} and \lwe{}\textemdash we run the reduction step described in \S\ref{subsec:data_method} (Stage 1.3) before performing model training and secret recovery.

\begin{table}[h!]
    \centering
    \small 
    \begin{tabular}{l|ccccc} \toprule
 {Dimension}        &  80 & 150 & 200 & 256 & 300 
 \\ \midrule 
   \tinyLWE{} max $h$     &  9 &  13 & 22 & 31 & 33
   \\
    \lwe{}  max $h$    &  9 & 12 & 21 & 32 & 32
    \\
    \bottomrule
    \end{tabular}
    \vspace{0.1cm}
    \caption{\small\textbf{\tinyLWE{} vs \lwe{}}. Values: highest $h$ recovered for each $n$.}
    \vspace{-0.5cm}
    \label{tab:tiny_vs_lwe}
\end{table}

There is little difference between the highest Hamming weight of recovered secrets for \tinyLWE{} and \lwe{} (Table~\ref{tab:tiny_vs_lwe}). In fact, \tinyLWE{} sometimes recovers larger Hamming weights than \lwe{}. Thus, we conclude that \tinyLWE{}, while greatly reducing the LWE samples needed for the attack, has no impact on performance. More detailed comparisons can be found in Table~\ref{tab:compare_TinyLWE_lwe} in Appendix~\ref{subsec:tiny_vs_lwe}.

\subsection{Resources needed for \system{}}
\label{subsec:cost}

The total cost of \system{} is the sum of the resources needed to preprocess data, train the model, and recover the secret.

\begin{table}[h!]
  \centering
  \small
\begin{tabular}{r|cccccc}
  \toprule
  n & 80  & 150 & 200 & 256 & 300 & 350 \\
  $\log_2 q$ & 7 & 13 & 17 & 23 & 27 & 32 \\ \midrule
  Cost per matrix & 0.01 & 3 & 16 & 52 & 106 & 194 \\
  (CPU.hrs) & & & & & \\
    Matrices needed & 34,800 & 14,600  & 10,800  & 8,300  & 7,100 & 6,000   \\ 
  \bottomrule
  \end{tabular}
  \vspace{0.1cm}
  \caption{\small\textbf{Resources needed for preprocessing.} Total resources needed to produce $2^{22}$ reduced samples is the work required for processing $2^{21}/ n$ matrices. Processing can be fully parallelized, so total time required is the number of CPU hours needed for one matrix. }
  \vspace{-0.5cm}
  \label{tab:bkz_preprocessing}
  \end{table}

\para{Data preprocessing} is the most resource intensive part of \system{}. To generate $2^{22}$ reduced samples, $2^{21}/n$ matrices must be reduced (one $n\times n$ matrix produces $2n$ reduced samples, see \S~\ref{subsec:data_method}).
As the dimension increases, the number of matrices needed scales down linearly. To avoid the exponential cost of BKZ-reduction \cite{BKZ}, we fix the block size to at most $\beta = 20$ so that the preprocessing step scales as a polynomial in $n$ and $\log q$. In practice, to save resources, we choose smaller $\beta$ for larger dimensions. Parameter choices for preprocessing are discussed further in \S\ref{subsec:ablate_preproc} below.

Table~\ref{tab:bkz_preprocessing} reports the preprocessing resources (in cpu$\cdot$hours) required for each~$n$. It is important to note that our preprocessing step is fully parallelizable.  Using as many CPUs as the number of matrices needed ($2^{21}/n$), the preprocessing step can be performed in the time required to reduce one matrix (e.g. $194$ hours for $n=350$). 

\para{Model training and secret recovery.}
The cost of training and recovery is proportional to the number of training epochs needed to recover the secret. 
Table~\ref{tab:cost} reports the average duration of one training epoch and associated secret recovery. All models use the same number of parameters, batch size ($128$) and epoch size ($2$ million examples), and are trained on one NVIDIA V100 GPU. 

Training time increases with dimension. This is expected, as the length of input sequences is $2n$, i.e. linear in the dimension, and training is slower on long sequences. For secret recovery, the time required for each method is dominated by the number of transformer inferences needed, multiplied by the time required for each inference.
The cross-attention method uses a constant number of inferences, direct recovery uses $15n$ inferences, and distinguisher recovery $200n$.
Like training, the time for a single inference scales linearly with $n$ because of increasing sequence length. Overall, cross-attention recovery scales linearly with $n$, and direct and distinguisher scale quadratically. 
In our experiments, secret recovery accounts for less than $10\%$ of the total time. We report the cost of training and recovery per epoch on one GPU, but both training and recovery time could be significantly reduced by parallelizing across many GPUs.

\begin{table}[h]
    \centering
    \small 
    \begin{tabular}{lcccccc} \toprule
    $n$             & 80 & 150 & 200 & 256 & 300 & 350 \\ 
    $\log q$        & 7  & 13  & 17  & 23  & 27  & 32  \\ \midrule
    Training        & 42 & 52 & 68 & 82  & 92  & 105 \\ 
    Secret Recovery & 1 & 2 & 3 & 5 & 7 & 8 \\ \bottomrule
    \end{tabular}
    \vspace{0.1cm}
    \caption{\small\textbf{Training and recovery time per epoch (minutes).} All models are trained on a single NVIDIA V100 GPU with batch size 128). 
    } 
    \vspace{-0.6cm}
    \label{tab:cost}
\end{table}

%% file: tables/table_success.tex
\begin{table}[t]
\small
  \centering
  \resizebox{0.47\textwidth}{!}{%
  \begin{tabular}{lccccccc} \toprule
 \multicolumn{2}{l}{ $n, \log_2 q$}  & \multicolumn{6}{c}{Hamming weight $h$}      \\ \midrule \midrule
  80, 7     & 4       & 5       & 6      & 7     & 8    & \bf 9    & 10    \\ \midrule
  success & 3/5     & 3/5     & 2/5    & 2/5    & 1/20     & 1/20  & 0/20 \\
  epoch & 2,5,6   & 0,1,4   & 0,3    & 0,8    & 3        & 4     &      \\ \midrule \midrule
  150, 13   & 9       & 10      & 11     & 12    & \bf 13   & 14   & 15    \\ \midrule
  success     & 4/5     & 2/5     & 3/5    & 1/5    & 1/20     & 0/20  & 0/20 \\
  epoch & 1,1,3,6 & 2,2     & 8,8,11 & 9      & 13       &       &      \\ \midrule \midrule
  200, 17 & 17      & 18      & 19     & 20     & 21       & \bf 22    & 23  \\ \midrule
  success       & 3/5     & 2/5     & 3/5    & 1/5    & 2/5      & 2/20  & 0/20 \\
  epoch & 1,1,8   & 2,11    & 2,3,9  & 7      & 7,10     & 12,17 &      \\ \midrule \midrule
  256, 23 & 26      & 27      & 28     & 29     & 30       & \bf 31    & 32   \\  \midrule
  success       & 4/5     & 1/5     & 1/5    & 3/5    & 3/5      & 4/20  & 0/20 \\
  epoch & 2,3,4,7 & 10      & 5      & 5,9,11 & 17,20,32       & 6,12,26,27  &      \\ \midrule \midrule
  300, 27 & 28      & 29      & 30     & 31    & 32   & \bf 33   & 34    \\  \midrule
  success       & 2/5     & 2/5     & 1/5    & 1/5    & 2/5      & 1/5   & 0/20 \\
  epoch & 6,7     & 6,13    & 11     & 11     & 21,31    & 39    &      \\ \midrule \midrule 
  350, 32 & 54   & 55      & 56      & 57     & 58    & 59  & \bf 60    \\ 
  \midrule
  success & 2/5   & 1/5 & 1/5 & 1/5 & 1/5 & 1/5 & 1/5 \\
  epoch & 10,20 & 10  & 46  & 42  & 39  & 18  & 38    \\\bottomrule
  \end{tabular}%
  }
  \vspace{0.1cm}
  \caption{\small\textbf{Success rates and number of epochs.} Highest recovered Hamming weight for each dimension $n$ is in bold.}
  \vspace{-0.4cm}
  \label{tab:success_epochs}
  \end{table}

%% file: ablation.tex
\section{Additional Results}
\label{sec:ablate}

Now, we consider the effect of different experimental choices on \system{}'s performance. This enables us to better understand the conditions under which \system{} succeeds or fails.

\subsection{Data preprocessing}
\label{subsec:ablate_preproc}
\begin{figure}[h]
\small
    \centering
    \includegraphics[width=0.45\textwidth]{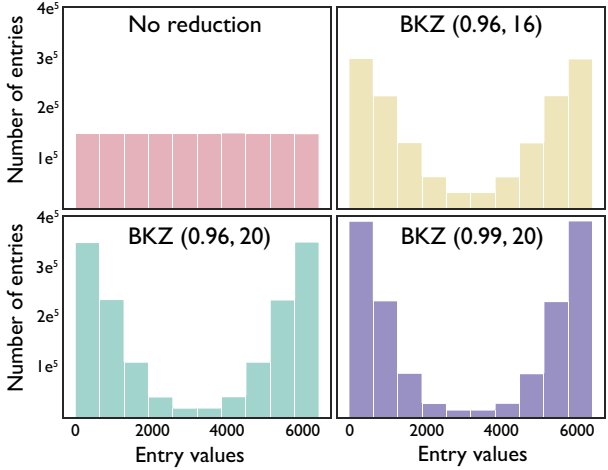}
    \vspace{-0.15cm}
    \caption{\small\textbf{Distribution of sample entry values as strength of norm reduction increases ($n=150$, $q=6421$)}. BKZ parameters: BKZ $(\beta, \delta)$. } 
    \vspace{-0.1cm}
    \label{fig:bkz_hist}
\end{figure}

\begin{figure}[h]
\small
    \centering
    \includegraphics[width=0.37\textwidth]{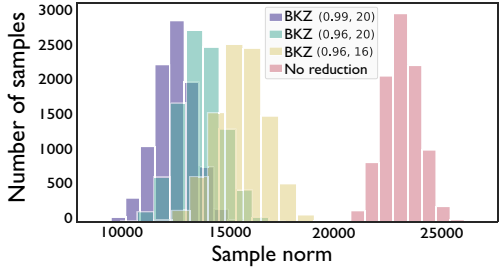}
    \vspace{-0.2cm}
    \caption{\small\textbf{Distribution of sample norms as strength of norm reduction increases ($n=150$)}. BKZ parameters listed as BKZ $(\beta, \delta)$. }
    \vspace{-0.2cm}
    \label{fig:bkz_norm}
\end{figure}

Through extensive experimentation, we observed that data preprocessing is critical to enabling the transformer to learn and allowing recovery of secrets with larger Hamming weight. Preprocessing changes the distribution of both the size of the entries of $\textbf{A}$ modulo~$q$ (shown for $n=150$ in Figure~\ref{fig:bkz_hist}) and the norm of its rows (see Figure~\ref{fig:bkz_norm}). Note that the goal of preprocessing is {\bf not} to obtain the shortest vector in the lattice like the classical uSVP, decoding, and dual attacks. Rather, its goal is to skew the distribution to make it more amenable to machine learning-based attacks.

\para{Choosing preprocessing parameters.} To determine the amount of preprocessing needed for optimal \system{} performance, we set various targets for the standard deviation of the entries of the rows of $\bf A$. 
For random matrices $\textbf A_{rand}$, the standard deviation of the entries is $\text{std}(\textbf A_{rand}) \approx q/\sqrt{12} \approx 0.29q$. After preprocessing, the standard deviation of the entries are smaller, e.g., for $n=256$, $\text{std}(\textbf A)\approx 0.1 q$. Empirically, we observed that reducing the ratio $\text{std}(\textbf A)/\text{std}(\textbf A_{rand})$ allows us to recover secrets with higher hamming weights.
In practice, we select parameters for our BKZ preprocessing step in \fplll{} by first processing a single matrix $\textbf{A}$ with various choices for ~$\beta$ and~$\delta$ and observing the resulting standard deviation. 
To prepare a whole dataset for training, we preprocess the $2^{21}/n$ matrices with parameters that reach a low standard deviation in a reasonable time;  when we did not recover secrets with the target $\approx 10\%$ density for $n=350$, we repeated the data generation with stronger parameters.

\para{Relationship between preprocessing and recovered $h$.} To examine how preprocessing parameters affect \system{}'s secret recovery, we ran four sets of experiments with values of $\beta$ and $\delta$ ranging from no preprocessing at all to the parameters used in \system{}.
Table~\ref{tab:bkz_vs_no} shows statistics of the data distribution and the largest $h$ recovered for these experiments. 
As preprocessing strength increases, for dimension $n=150$  (middle columns), the weight $h$ of recovered secrets increases up to $12$. As shown in the last two columns of Table~\ref{tab:bkz_vs_no}, for $n=350$, stronger preprocessing parameters decrease the standard deviation of the entries and enable recovery of larger weight $h$ secrets (up to $h=60$).

\input{tables/table_bkz.tex}

\para{Relationship between preprocessing and $n$.} Table~\ref{tab:norm_reduction} presents the highest Hamming weight recovered and standard deviations of reduced $\textbf A$ for different $n$. We observe that, even though we decreased the block size and $\delta$ to reduce preprocessing time for larger dimensions ($n=256, 300$ and $350$), as long as the standard deviation of the entries is low enough, \system{} recovers secrets with $> 10\%$ sparsity.

\begin{table}[t]
\small
    \centering
    \begin{tabular}{cc|cccc}
        \toprule
        $ n$ & $ \log_2 q$ & $\delta$ & $\beta$ &  std(\textbf A)$/\text{std}(\textbf A_{rand})$ & max $h$  \\ \midrule
80  & 7  & 0.99 & 20 & 0.78 & 9  \\
150 & 13 & 0.99 & 20 & 0.53 & 13 \\
200 & 17 & 0.99 & 20 & 0.40 & 22 \\
256 & 23 & 0.96 & 18 & 0.33 & 31 \\
300 & 27 & 0.96 & 16 & 0.32 & 33 \\
350 & 32 & 0.96 & 14 & 0.25 & 60 \\ \bottomrule
    \end{tabular}
    \vspace{0.1cm}
\caption{\small\textbf{Preprocessing parameters for BKZ in \fplll{} for \system{}'s best secret recoveries}. $\beta:$ block size, 
$\delta:$ LLL\_DELTA.} 
\vspace{-0.6cm}
    \label{tab:norm_reduction}
\end{table}

\subsection{Encoding base}
\label{subsec:base_ablation}

We explore how $B$, the base used to encode $b$ and the coordinates of~$\textbf a$ (\S\ref{subsec:transformers}) during model training, affects \system{} performance. Table~\ref{tab:base_n150_ablation} presents the impact of different base choices on secret recovery for $n=150$, $q=6421$. For a small modulus like this, no buckets are needed, i.e. $r=1$. To keep input sequences short, all integers modulo $q$ should be encoded in two tokens, i.e. $B\geq \sqrt q$. However, values of~$B$ close to $\sqrt q$ result in worse secret recovery. Recovery rates are highest when $B = \lfloor q/k \rfloor$ with $k=6$ or $8$. 

\begin{table}[t]
\small
    \centering
    \begin{tabular}{c|cccc} \toprule
    base & $h=9$ & $h=10$ & $h=11$ & $h=12$\\\midrule
$81 \approx \sqrt{q}$    & 1/5 & 0/5 & 0/5 & 0/5 \\
$402 \approx q/16$ & 3/5 & 2/5 & 0/5 & 0/5\\
$803 \approx q/8$ & 3/5 & 2/5 & 2/5 & 2/5\\
$\mathbf{1071 \approx q/6}$ & \bf{4/5} &  \bf{2/5} &  \bf{3/5} & \bf{1/5}\\
$1606 \approx q/4$ & 3/5 & 2/5 & 0/5 & 0/5  \\ \bottomrule
    \end{tabular}
    \vspace{0.1cm}
    \caption{\small\textbf{Secret recovery rate for different bases $B$.} $n=150$, $q=6421$. \system{} parameters and results are in bold.} 
    \label{tab:base_n150_ablation}
    \vspace{-0.6cm}
\end{table}

Table~\ref{tab:base_n256_ablation} presents a similar study of base $B$ and bucket size $r$ for larger $n=256$ and $q=6139999$. When $q$ is this large, using base $B=\lfloor q/k \rfloor$ would result in a vocabulary that is too large for the transformer to learn efficiently. Hence, we tokenize the less significant digits in buckets of size~$r$, as described in \S\ref{subsec:transformers}. Small values of $B\approx \sqrt q$ and large values $B=q/4$ do not seem to result in good secret recovery. For $B=q/8$ and $q/16$, bucket sizes $r=128$ and $r=512$ have comparable performance.

\begin{table}[t]
    \centering
    \small
    \begin{tabular}{cc|cccc} \toprule
        \multicolumn{2}{c|}{Encoding} & \multicolumn{4}{c}{Hamming weight $h$}       \\ 
        base $B$                & $r$    & 26 & 27 & 28 & 29 \\ \midrule 
        $2478\approx \sqrt{q}$                         & 1             & 0/5 & 0/5 & 0/5 & 0/5 \\ \midrule
        \multirow{3}{*}{$383750\approx q/16$}      & 32            & 4/5 & 1/5 & 1/5 & 0/5 \\
                                     & 128           & 3/5 & 2/5 & 1/5 & 3/5 \\
                                     & 512           & 4/5 & 2/5 & 1/5 & 3/5 \\ \midrule
        \multirow{3}{*}{ $\mathbf{767500\approx q/8}$}      & 32        & 4/5 & 1/5 & 1/5 & 1/5 \\
                                     & \bf 128           & \bf 4/5         & \bf 1/5         & \bf  1/5         & \bf 3/5         \\
                                     & 512           & 4/5 & 2/5 & 1/5 & 3/5 \\ \midrule
        \multirow{3}{*}{$1535000\approx q/4$}     & 32            & 0/5         & 0/5         & 0/5         & 0/5         \\
                                     & 128           & 0/5         & 0/5         & 0/5         & 0/5         \\
                                     & 512           & 1/5         & 0/5         & 0/5         & 0/5     \\ \bottomrule   
        \end{tabular}%
        \vspace{0.1cm}
    \caption{\small\textbf{Secret recovery rates for different bases $B$ and bucket sizes $r$.} $n=256$, $q=6139999$. \system{} parameters are in bold. }
    \vspace{-0.8cm}
    \label{tab:base_n256_ablation}
\end{table}

\subsection{Model architecture}
\label{subsec:model_ablate}

All \system{} experiments use the same model architecture (see \S~\ref{subsec:transformers} for details). However, \salsa{} reported improved performance for larger $n$ with larger models, specifically increased embedding dimensions. Also, the number of attention heads used in \system{}, $4$ in the encoder and decoder, is low, compared to common transformer architectures. Most transformers with $512$ dimensions use $8$ heads. Thus, we explore the impact of larger dimensions and number of heads in the encoder and decoder, on secret recovery (Table~\ref{tab:model_size2}) for $n=350$. As the table shows, increasing dimension and heads do {\em not} result in better performance. This contrasts with results in NLP, where performance usually increases with model size. We believe this is because the LWE problem differs from traditional NLP tasks.

\begin{table}[h!]
    \small
    \centering
    \begin{tabular}{c|ccccc} \toprule
    \multirow{3}{*}{\begin{tabular}[c]{@{}c@{}} embedding size \\ encoder/decoder \\ \end{tabular}} & \multicolumn{5}{c}{number of attention heads} \\ & \multicolumn{5}{c}{encoder/decoder/cross attention}\\ \cmidrule{2-6}
                         & {\bf 4/4/4} & 4/4/8 & 4/4/16 & 8/8/8 & 8/8/16 \\ \midrule 
    {\bf 1024 / 512} & {\bf 60}           & 58           & 58            & -            & -             \\
    1024 / 768 & -            & 58           & 60            & 57           & 55            \\
    1280 / 512 & -            & 60           & 58            & 58           & 58         \\ \bottomrule
    \end{tabular}%
    \vspace{0.1cm}
    \caption{\small\textbf{Effect of architecture on \system{}'s performance.} Data shown is the highest Hamming weight recovered for $n=350$. \system{}'s parameters are in {\bf bold}. }
    \vspace{-0.8cm}
    \label{tab:model_size2}
    \end{table}

\subsection{Model initialization}
\label{subsec:seeds_ablate}

Transformer parameters are randomly initialized before training, and these initial values may impact the performance. This is known as the ``lottery ticket'' phenomenon: models sometimes learn better, or faster, with  different initial parameter values. We explore this effect in $4$ experiments for $n=200$ and $h=19$. Each experiment in Table~\ref{tab:seed_ablate} has a different secret; for each secret we train $20$ transformers, each initialized with a different seed. In experiment 1, the secret is recovered for all $20$ seeds, at epoch 2 to~7. For experiments 2 and 3, the secret is recovered about 3/4 of the time, between 6 and 25 epochs. In experiment 4, the secret is never recovered.

\begin{table}[t]
    \small
    \centering
    \begin{tabular}{c|ccc} \toprule
     Experiment & Success  & Mean epoch & Min, max epochs \\ \midrule
    1   &   20/20 &  4.2  & 2, 7          \\
    2      &   12/20  & 12.1 & 8, 25         \\
    3   &   15/20 &  9.1  & 6, 16         \\
    4     &   0/20 & - & -       \\  \bottomrule
    \end{tabular}
    \vspace{0.1cm}
    \caption{\small\textbf{Effect of model initialization on secret recovery.} $n=200$, $h=19$.} 
    \vspace{-0.9cm}
    \label{tab:seed_ablate}
\end{table}

This sheds light on results from \S\ref{subsec:best_results}. There, we observed significant variance in the number of epochs needed for secret recovery, for given $n$ and $h$. Initialization seems to be an important factor and suggests a possible improvement to \system{} when significant compute resources are available. By training several transformers with different initializations on the same data, \system{}'s chances of secret recovery improve, as does training speed\textemdash training can stop for all models once one recovers the secret.

\subsection{Secret recovery methods}
\label{subsec:secret_ablate}

\system{} leverages $4$ secret recovery methods (\S~\ref{subsec:recovery}): distinguisher, direct, cross-attention, and combined. The first three methods output {\it bit scores} and secret guesses $\mathbf{s'}$. The {\it bit scores} rank the likelihood of individual secret bits having value $1$. The  \textit{combined secret recovery} method allows \system{} to create additional secret guesses by aggregating the scores of the previous methods. Table~\ref{tab:secret_recovery} reports the successes/attempts of all secret recovery methods for fixed $n/q$ and varying $h$.
We only report the method(s) that succeed first: we terminate each experiment after successful recovery.
We say that the {\it combined} method is successful if and only if it recovered the secret when no individual method could. 
If an individual method succeeds, the combined method typically succeeds as well.

\begin{table}[h!]
  \centering
  \small
  \begin{tabular}{l|ccccccc} \toprule
  $n, \log_2 q$  & \multicolumn{7}{c}{Hamming weight $h$}      \\ \midrule \midrule
  80, 7           & 4   & 5   & 6   & 7   & 8    & 9    & 10   \\ \midrule
success         & 3/5 & 3/5 & 2/5 & 2/5 & 1/20 & 1/20 & 0/20 \\
Distinguisher   & 3/5 & 3/5 & 2/5 & 2/5 & 1/20 & 1/20 & 0/20 \\
Direct          & 2/5 & 1/5 & 1/5 & 1/5 & 1/20 & 1/20 & 0/20 \\
Cross-attention & 3/5 & 1/5 & 2/5 & 0/5 & 1/20 & 0/20 & 0/20 \\
Combined        & 0/5 & 0/5 & 0/5 & 0/5 & 0/20 & 0/20 & 0/20 \\ \midrule \midrule
150, 13         & 9   & 10  & 11  & 12  & 13   & 14   & 15   \\ \midrule
success         & 4/5 & 2/5 & 3/5 & 1/5 & 1/20 & 0/20 & 0/20 \\
Distinguisher   & 2/5 & 1/5 & 2/5 & 0/5 & 0/20 & 0/20 & 0/20 \\
Direct          & 2/5 & 0/5 & 0/5 & 0/5 & 1/20 & 0/20 & 0/20 \\
Cross-attention & 1/5 & 0/5 & 1/5 & 1/5 & 0/20 & 0/20 & 0/20 \\
Combined        & 0/5 & 1/5 & 0/5 & 0/5 & 0/20 & 0/20 & 0/20 \\ \midrule \midrule
200,17          & 17  & 18  & 19  & 20  & 21   & 22   & 23   \\ \midrule
success         & 3/5 & 2/5 & 3/5 & 1/5 & 2/5  & 2/20 & 0/20 \\
Distinguisher   & 2/5 & 1/5 & 2/5 & 1/5 & 0/5  & 0/20 & 0/20 \\
Direct          & 0/5 & 0/5 & 1/5 & 0/5 & 0/5  & 0/20 & 0/20 \\
Cross-attention & 1/5 & 0/5 & 0/5 & 0/5 & 0/5  & 1/20 & 0/20 \\
Combined        & 0/5 & 1/5 & 1/5 & 0/5 & 2/5  & 1/20 & 0/20 \\ \midrule \midrule
256, 23         & 26  & 27  & 28  & 29  & 30   & 31   & 32   \\ \midrule
success         & 4/5 & 1/5 & 1/5 & 3/5 & 3/5  & 4/20 & 0/20 \\
Distinguisher   & 3/5 & 1/5 & 1/5 & 1/5 & 1/5  & 0/20 & 0/20 \\
Direct          & 0/5 & 0/5 & 0/5 & 0/5 & 0/5  & 0/20  & 0/20  \\
Cross-attention & 2/5 & 0/5 & 1/5 & 3/5 & 2/5  & 2/20 & 0/20 \\
Combined        & 1/5 & 0/5 & 0/5 & 0/5 & 0/5  & 2/20 & 0/20 \\ \midrule \midrule
300, 27         & 28  & 29  & 30  & 31  & 32   & 33   & 34   \\ \midrule
success         & 2/5 & 2/5 & 1/5 & 1/5 & 2/5  & 1/5  & 0/20 \\
Distinguisher   & 2/5 & 0/5 & 0/5 & 0/5 & 0/5  & 0/5  & 0/20 \\
Direct          & 0/5 & 0/5 & 0/5 & 0/5 & 0/5  & 0/5  & 0/20 \\
Cross-attention & 1/5 & 2/5 & 1/5 & 0/5 & 1/5  & 1/5  & 0/20 \\
Combined        & 0/5 & 0/5 & 0/5 & 1/5 & 1/5  & 0/5  & 0/20 \\ \midrule \midrule
350, 32         & 54  & 55  & 56  & 57  & 58   & 59  & 60   \\ \midrule
success         & 2/5   & 1/5 & 1/5 & 1/5 & 1/5 & 1/5 & 1/5 \\
Distinguisher   & 0/5   & 0/5 & 0/5 & 0/5 & 1/5 & 0/5 & 0/5 \\
Direct          & 0/5   & 0/5 & 0/5 & 0/5 & 0/5 & 0/5 & 0/5 \\
Cross-attention & 0/5   & 1/5 & 0/5 & 0/5 & 0/5 & 0/5 & 1/5 \\
Combined        & 2/5   & 0/5 & 1/5 & 1/5 & 0/5 & 1/5 & 0/5 \\ \bottomrule
  \end{tabular}%
  \vspace{0.1cm}
  \caption{\small\textbf{Secret recovery successes/attempts for \system{}'s four recovery methods.} For each ($n$, $q$, $h$) setting, we report the number of secrets \system{} recovers out of attempted attacks (``success'' row) and the number of recoveries by each individual method (``Distinguisher'' through ``Combined'' rows). If two methods succeed in the same training epoch, we report both successes, so individual recoveries may exceed the number of total successes. \rev{Combined method successes are only reported when all other methods fail.}
  }
  \vspace{-0.8cm}
  \label{tab:secret_recovery}
  \end{table}

\looseness=-1 Two trends are evident in Table~\ref{tab:secret_recovery}. First, the direct recovery method is outperformed by other methods as $n$ increases. 
It works well at $n=80$, but for $n \ge 256$, it is either slower than other methods or fails to recover the secret.
Recall that direct recovery works when for every bit $i$ of the secret, the model prediction $\T(K \cdot \mathbf{e}_i)$ 
corresponds to the secret bit:  large when the secret bit is 1 and small otherwise.
This happens with lower probability as $n$ grows. 
Second, the combined recovery method performs better as~$n$ increases.
Probably for larger $n$, 
individual methods cannot glean information about all secret bits, but each gains some information about some bits. Thus, combining their scores may allow additional recoveries. 

%% file: tables/table_bkz.tex
\begin{table}[h!]
    \centering
    \small
    \begin{tabular}{l|cccc|cc}
    \toprule
    $\hspace{18mm}n,\log_2 q$ & \multicolumn{4}{c}{150, 13} & \multicolumn{2}{c}{350, 32}\\
    $\hspace{22mm}\delta$ & - & 0.96 & 0.96 & 0.99 & 0.93 & 0.96\\
    $\hspace{22mm}\beta$ & - & 16 & 20 & 20 & 14 & 14\\
    \midrule
    \bf{highest $h$}  & \bf - & \bf 5 & \bf 8 & \bf 12 & \bf 25 & \bf 60\\
    std(\textbf A)$/\text{std}(\textbf A_{rand})$  & 1 & 0.667 & 0.578 & 0.526 & 0.331 & 0.253\\
    norm(\textbf A)$/\text{norm}(\textbf A_{rand})$   & 1 & 0.669 & 0.581 & 0.528 & 0.332 & 0.253\\
    cost / matrix (hours)   & 0 & 0.5 & 0.9 & 3.1 & 152 & 194 \\ 
    time out (hours)   & - & 1 & 2 & 5.5 & - & -\\ \bottomrule
    \end{tabular}%
    \vspace{0.1cm}
    \caption{\small\textbf{Highest weight $h$ secret recovered for varying $\delta$ and/or $\beta$ ($n=150,350$)}.  $\text{std}({\bf A})$: standard deviation of ${\bf A}$'s coefficients post-reduction; $\text{norm}({\bf A})$: average norm of $\textbf{A}$'s rows post-reduction.}
    \vspace{-0.6cm}
    \label{tab:bkz_vs_no}
\end{table}

%% file: comparison.tex
\section{Comparison to existing LWE attacks}
\label{sec:compare}

Finally, we compare \system{}'s performance against classical lattice attacks. This is a difficult task, given both the significant differences in methodology between \system{} and existing attacks, as well as the lack of reported concrete running times for existing attacks. In practice, the training stage of \system{} takes less time than pre-processing the data (see Table~\ref{tab:bkz_preprocessing} and Table~\ref{tab:cost} in \S~\ref{subsec:cost}),
so we focus on comparing the cost of preprocessing with the cost of {\it classical lattice reduction attacks} such as uSVP, decoding, and dual attacks. 
As \system{} uses the \fplll{} package for lattice reduction algorithms, we compare the running times of \system{} with the uSVP attack, run using \fplll{}.

The LWE Estimator software package~\cite{LWEestimator} is used to estimate the cost of classical lattice reduction attacks. The LWE Estimator uses theoretical formulas and heuristic estimates to predict which block size will be required for BKZ to recover the secret for a given lattice parameter size.  These estimates are widely used to set parameters and estimate security at parameter sizes for which it is impossible to actually run these classical attacks (they would not terminate in our lifetimes). 
Concrete running times for actual successful attacks can be found in a few places in the literature, e.g. in~\cite{CCLS, laine2015key, albrecht2017revisiting,bai2019refined}, and we find those useful for comparison here. 
In particular, for dimensions $n \leq 200$, we chose values of $\log q$ strictly smaller than those used in \cite{CCLS}; for dimensions $n=256, 300 $ and $350$, we use much smaller values of $\log q$ than  \cite{laine2015key}: for instance, for $n=350$, we use $\log q = 32$, much smaller than the value $\log q = 52$ in \cite{laine2015key}. 

We present here 3 ways to quantify, estimate, and compare with pure lattice reduction attacks: the LWE estimator, concrete timings for running the uSVP attack at small sizes, and theoretical and heuristic formulas.

\para{LWE Estimator.} Table~\ref{tab:estimator} presents the block size and estimated cost 
for classical attacks, to compare against \system{}'s successful secret recoveries (using the highest $h$ achieved by \system{}).
LWE Estimator \cite{LWEestimator} costs are listed in terms of the number of operations in~$\ZZ_q$, the cost of which can be approximated by $(\log q)^2$. For example, for $n=256$, the LWE estimator predicts that the uSVP attack should succeed with block size $40$ and cost about $2^{41.8}$ operations in $\ZZ_q$. \system{} uses block size~$18$.

\begin{table}[t]
\centering
\small
\begin{tabular}{cc|lcc}\toprule 
$n$ & $q$ & \text{best attack} & \text{cost} & \text{block size}  \\
\midrule
$80$ &	$113$ &  BDD  & $2^{48.0}$ &  $\beta = 63$ \\	
$150$ & 	$6421$ & BDD & $2^{42.7}$ &   $\beta = 44$	 \\				
$200$ & 	$130769$  & BDD & $2^{41.8}$ & $\beta = 41$\\
$256$ &  $6139999$ & uSVP/BDD & $2^{41.8}$ & $\beta = 40$\\ 
$300$  & $94056013$ & uSVP &  $2^{41.9}$  &  $\beta= 40$ \\
$350 $  & 	$3831165139$  & uSVP &  $  2^{42.0}$ &   $\beta= 40$ \\
\bottomrule
\end{tabular}
\vspace{0.1cm}
\caption{\small\textbf{LWE Estimator \cite{LWEestimator} estimates} for \system{}'s most successful recoveries (see Table~\ref{tab:best_results}). Cost: number of operations in $\ZZ_q$.}
\label{tab:estimator} 
\vspace{-1.0cm}
\end{table}

\begin{table*}[]
    \centering
    \small
\begin{tabular}{cc|ccccc|cclc} \toprule
\multirow{4}{*}{$n$, $\log_2q$} & \multirow{4}{*}{$h$} & \multicolumn{5}{c|}{PICANTE}& \multicolumn{4}{c}{uSVP attack with BKZ 2.0 and early-abort }\\
&  &  &   &  \multicolumn{2}{c}{\em Preprocessing}  & {\em Training}  &  \multicolumn{4}{c}{}\\ 
& & $\beta$ & success & CPU.hrs & $\#$ matrices& CPU.hrs                              & $\beta$ & success & success time & fail time\\
& & & & per matrix & & per epoch & & &(CPU.hrs) &  (CPU.hrs)
\\
\midrule
\multirow{4}{*}{80, 7}                        & \multirow{2}{*}{6, 7}                        & \multirow{2}{*}{20}   & \multirow{2}{*}{4/10}       & \multirow{2}{*}{0.01} & \multirow{2}{*}{34800}  & \multirow{2}{*}{0.7}    & 60                    & 2/10                        & 8, 12                              & 7.8                           \\
                                              &                                               &                       &                             &                                                  &                   &                & 65                    & 6/10                        & 9, 10, 14, 18, 40, 85                            & 72.1                          \\ \cmidrule{2-11}
                                              & \multirow{2}{*}{8, 9}                        & \multirow{2}{*}{20}   & \multirow{2}{*}{2/40}       &  \multirow{2}{*}{0.01} & \multirow{2}{*}{34800}  & \multirow{2}{*}{0.7 }    & 55                    & 0/10                        & \textemdash                              & 2.4                           \\
                                              &                                               &                       &                             &                                                  &                     &              & 60                    & 1/10                        & 12                              & 6.9                           \\ \midrule
\multirow{4}{*}{150, 13}                      & \multirow{2}{*}{9,10}                        & \multirow{2}{*}{20}   & \multirow{2}{*}{6/10}       &  \multirow{2}{*}{3.1}  & \multirow{2}{*}{14600}   & \multirow{2}{*}{0.9 }    & 50                    & 5/10                        & 26, 30, 31, 35, 35                             & 22.9                          \\
                                              &                                               &                       &                             &                                                  &                   &                & 55                    & 8/10                         & 19, 19, 23, 23, 23, 23, 28, 28                            & 22.7                           \\ \cmidrule{2-11}
                                              & \multirow{2}{*}{11, 12}                        & \multirow{2}{*}{20}   & \multirow{2}{*}{4/10}       & \multirow{2}{*}{3.1} &  \multirow{2}{*}{14600} & \multirow{2}{*}{0.9 }    & 50                    & 2/10                        & 22, 39                             & 9.5                          \\
                                              &                                               &                       &                             &                                                  &                   &                & 55                    & 4/10                        & 14, 19, 24, 33                             & 5.6                           \\ \midrule
\multirow{2}{*}{200, 17}                      & 18, 19                        & 20   & 5/10       & 16 &  10800  & 1.2    & 45                    & 6/10                        & 12, 13, 18, 21, 21, 21                             & 7.7                                              \\ \cmidrule{2-11}
                                              & 20, 21                                         & 20                    & 3/10                        & 16  & 10800                   & 1.2                    & 45                    & 3/10                        & 13, 21, 25                             & 12.9                           \\ \midrule
\multirow{3}{*}{256, 23}                      & 26, 27                                         & 18                    & 5/10                        & 52  & 8300                  & 1.5                 & 40                    & 4/10                         & 203, 221, 243, 265                            & 189.1                            \\ \cmidrule{2-11}
                                              & 28, 29                                         & 18                    & 4/10                        & 52 & 8300                    & 1.5                    & 35                    & 7/10                         & 238, 246, 249, 269, 284, 303, 348                                & 241.9                          \\ \cmidrule{2-11}
                                              & 30, 31                                         & 18                    & 4/10                        & 52  & 8300                    & 1.5              & 35                    & 5/10                         & 231, 255, 263, 330, 336                            & 171.7                       \\ \bottomrule 
\end{tabular}
\vspace{0.1cm}
\caption{\small\textbf{Concrete running times (CPU.hrs) for uSVP attacks and corresponding \system{} costs.} Training cost includes secret recovery time. 
\system{} uses BKZ, the uSVP attack uses BKZ2.0 \cite{CN11_BKZ}, see the discussion in \S~\ref{subsec:data_method}. 
In all uSVP attacks, we use $\omega = round(\sqrt{2}\sigma) = 4 $. 
Legend: CPU.hrs: CPU hours. fail time: average time for the failed experiments.
}
\vspace{-0.7cm}
\label{tab:concrete_time}
\end{table*}

\para{Concrete running times.}
Table~\ref{tab:concrete_time} presents concrete running times for the following attack:
We run the primal uSVP attack, using Kannan's embedding and BKZ2.0 \cite{CN11_BKZ} with different block sizes. The dimension for the Kannan's embedding is determined as in \cite{CCLS}.
We choose block sizes close to the block size predicted by the LWE Estimator and compare \system{} against attacks with similar success probability.  We ran the classical attacks for dimensions up to $n=256$ with the block size predicted by the LWE Estimator ($\beta =40$). 
For $n=300, 350$, already the first loop in BKZ takes longer than $3$ days; the full attack was taking too long to run. 

The uSVP attack was run using the \fplll{} package on the same machine as the norm-reduction step of \system{}.  We did not use any optimization for either of the attacks. 
We see that for $n=80$ and $h=9$, \system{} with $\beta =20$ achieves similar success to uSVP with $\beta= 60$. The uSVP attack takes about 10 hours to succeed; \rev{for this $n=80$, $h=9$ setting,} the time spent on data preprocessing for \system{} is negligible with enough parallelization and the training (run on 1 GPU) for successful recoveries took $5$ epochs of about $0.7$ hours each. The time spent by \system{} is therefore about 4 hours. 
For $n=256$ and $h=31$, the uSVP attack with block size $\beta =35$ (smaller than predicted by the LWE estimator) took a minimum of $231$ hours to succeed;  \system{} with sufficient parallelization needs $52$ hours for data pre-processing and a minimum of $10$ hours for training, so the total time is about $62$ hours.

These timings are rough estimates. Optimizations to lattice-reduction for the uSVP attack could also speed up the data preprocessing of \system{}. We did not include any possible savings from parallelizing training and secret recovery methods (see \S\ref{subsec:cost}).

\para{Theoretical analysis.} Denote by $BKZ(d, \beta)$ 
the (classical) cost of BKZ reduction in dimension $d$ with block size $\beta$. It can be estimated~\cite{HES} as $2^{0.292\beta + c} << BKZ(d, \beta) < 8d \cdot 2^{0.292 \beta + c}$,
where the cost $SVP(\beta) = 2^{0.292\beta+ c}$ is the cost of the SVP oracle in dimension~$\beta$ (a major step in the BKZ-reduction algorithm). 
The constant $c $ depends on the attack model\textemdash $16.4$ for sieving and $0$ for others. The upper bound arises from the estimated $8$ runs (full loops) of the BKZ-reduction, and hence $8d$ $SVP(\beta)$ oracle calls, needed.

The uSVP attack solves the shortest vector problem in dimension~$d > n$; 
\system{} applies the BKZ-reduction to lattices of dimension~$d= 2n$ but
keeps the block size close to constant ($\beta = 20$, and decreases the block size in larger dimensions for efficiency). We choose this because the cost of BKZ reduction scales exponentially with the block size. While we do not know if we can use constant block size $\beta = 20$ for all dimensions, we expect our block size to grow slower than block sizes required for lattice-reduction attacks.

\para{High level comparison.}
\system{} compares with classical lattice reduction attacks as follows: 
\system{} succeeds in recovering the secret vector {\it using much smaller block size than pure lattice reduction attacks, at the expense of processing many more matrices (2.2 million/n matrices)}.  
Because this step is run in parallel, \system{} recovers secrets faster than the uSVP attack but uses many more CPUs for parallel processing. As the dimension increases and/or $\log q$ decreases, we expect the advantage of \system{} to grow, 
due to the exponential cost of the lattice reduction attacks based on BKZ.
Future work may produce a more efficient way to preprocess the data or reduce the amount of data needed for training.

%% file: discussion.tex
\vspace{-0.3cm}
\section{Discussion}
\label{sec:discuss}

Our attack, \system{}, demonstrates a dramatic improvement over \salsa{}, the only prior work on attacking LWE with Machine Learning. \salsa{} pioneered the use of ML models in cryptanalysis of LWE, but only recovered secrets for small LWE problems. 
In contrast, \system{} successfully recovers LWE binary secrets with sparsity up to $10\%$, for dimensions up to $350$. It does so using only $4n$ LWE samples, a realistic assumption in practice. \system{}'s performance is competitive with that of known state-of-the-art attacks on LWE, particularly when sufficient compute resources are available, as \system{}'s novel data preprocessing step can be parallellized.

\para{Mastermind.} One way to think about the role of our novel preprocessing step is in analogy with the game Mastermind.  In Mastermind, a secret made up of 4 pegs of 6 possible colors is hidden from the guesser. The guesser makes queries of 4 pegs of different colors, and query responses indicate how many pegs matched the color and/or position of secret pegs.  Binary secret LWE can be thought of as Mastermind with $n$ positions and $2$ colors, ignoring error.  

In \system{}, the trained model serves as an engine for answering queries about the secret. Consider two extreme types of queries.  If you submit a vector with all entries constant, $(f,f,...,f)$, (i.e. very low entropy), 
you only get the Hamming weight\textemdash no information about the position of the $1$s.  On the other hand, the Direct secret recovery approach makes queries of the form $(0,...,0,K_i,0,...,0)$, which gives information only about the $i^{th}$ bit of the secret.  Submitting queries with random entries (maximal entropy) does not clearly give any particular type of information.

\system{}'s preprocessing step reduces the entropy of LWE samples, making it more likely that queries such as those in the Direct secret recovery method bear some similarity to the training samples. So in some sense our approach is ML for Mastermind (or Wordle).

\para{Scaling to larger Hamming weights.} \rev{ \salsa{} only recovered secrets with Hamming weights $h=3$ or $4$. \system{} recovers larger $h$ (up to $31$ for $n=256$ and $60$ for $n=350$), but recovering even larger $h$ (general binary secrets) is an important challenge for future work. As it stands, the \system{} attack can be countered by using general binary secrets. One way to scale to larger $h$ is to improve the preprocessing step. The more the variance of the training set's coordinates is reduced, the higher $h$ secrets \system{} recovers.}

\rev{
Our intuition regarding the relationship between preprocessing and recoverable $h$ is as follows (see~\cite{li2023salsa} for a detailed analysis). Given a secret $\textbf s$ with Hamming weight $h$ and dimension $n$, and a vector $\textbf a$ with coordinates uniformly sampled over $(-q/2,q/2)$, the dot product $\textbf a\cdot \textbf s$ is a sum of $h$ uniform random variables with mean $\mu = 0$ and standard deviation $\sigma = q/\sqrt{12}$. Thus, as $h$ grows, the distribution of $\textbf a\cdot \textbf s$ is roughly normal with $\mu = 0$ and $\sigma = nq \sqrt {h /12}$.
Thus, in $68\%$ of cases, the value of $\textbf a\cdot \textbf s$ will remain within one $\sigma$ of $\mu=0$, i.e. span a range of $2q \sqrt {h/12}=q \sqrt{h/3}$. If $h=3$, this range is $q$: $\textbf a\cdot \textbf s$ spans only one period of the modulus. This explains why \salsa{} has difficulty recovering secrets with $h \ge 4$: the model must learn modulus wrapping. Let $\alpha$ be the 
reduction factor 
$\textbf{std}(\textbf A)/\text{std}(\textbf A_{rand})$ 
achieved via preprocessing (see  Section~\ref{subsec:ablate_preproc} and Table~\ref{tab:bkz_vs_no}). Then $\textbf a\cdot \textbf s$ is a random variable with mean $\mu$=$0$, 
and $\sigma = \alpha q \sqrt{h/12}$. If $h<3/\alpha^2$, then $\textbf a \cdot \textbf s$ will span only one modulus, enabling easier learning.
This suggests that larger $h$ may be recovered by improving preprocessing.} 

\para{Ethical considerations.} Although \system{} demonstrates significant progress towards attacking real-world LWE problems with sparse binary secrets, it cannot yet break problems with real-world-size parameters.  In particular, the LWE schemes standardized by NIST use smaller modulus $q$ and non-sparse secret distributions. Hence, we do not believe our paper raises any ethical concerns. Nonetheless, we shared a copy of the current paper with the NIST Cryptography group, to inform them of our approach.

\para{Future directions.} More work is needed to better understand the effect of the data preprocessing step, since we observe that we only need a $5\%$ reduction of data entropy to succeed (Table~\ref{tab:bkz_vs_no}). Additionally, there may be better ways to preprocess the data to improve transformer learning, which are less costly than using BKZ.
In the future, the model training and secret recovery components of the attack could benefit from parallel runs across multiple GPUs, given our observation that different transformer initializations may result in different speeds of secret recovery (\S~\ref{subsec:seeds_ablate}). Furthermore, improvements to transformer architecture and secret recovery methods may enable recovery of secrets with more complex parameter settings. \rev{In particular, future work could explore the use of simpler model architectures to reduce memory and time costs. For example, prior work shows RNNs and LSTMs can perform modular addition, but~\cite{palamasinvestigating} suggests that this is difficult for FFNs. Finally, cross-attention secret recovery suggests that useful information can be gleaned from inspecting models' intermediate representations. Better understanding of these would be interesting future work.
}  

\vspace{-0.1cm}
\section*{Acknowledgements}
We thank Mark Tygert and Matteo Pirotta for many helpful discussions, and the anonymous reviewers for suggestions and edits.

%% file: appendix.tex
\section{Appendix}

\subsection{Comparison of \tinyLWE{} and LWE}
\label{subsec:tiny_vs_lwe}

Table~\ref{tab:compare_TinyLWE_lwe} compares secret recovery performance of models trained using samples generated via \system{}'s TinyLWE approach ($4n$ initial samples) vs. a baseline approach ($2^{22}$ initial samples).

\input{tables/table_tinyvslwe.tex}

\subsection{Statistical properties of secret verification}\label{app:verif}

At the end of the secret recovery phase, we are provided a secret guess $\textbf s'$, that we need to check. To do so, we use the original $m=4n$ LWE samples $(\textbf a_i,b_i)$, and compute the~$m$ residuals $r_i=b_i- \textbf a_i\cdot \textbf s'$. If the secret is recovered, we expect the $r_i$ to have the same standard deviation as a LWE sample, i.e. $\sigma$. Otherwise, we expect the standard deviation to be that of the uniform distribution, i.e. $q/\sqrt {12}$. The standard deviation of residuals is estimated by the formula: 
\[
\sigma_{\text{emp}} = \sqrt{\frac{1}{m-1}\sum_{i=0}^{m-1}(r_i-\overline{r})^2}
\]

\noindent Lower and upper confidence intervals, with level $100(1-\alpha)\%$ are:
\[\sigma_{emp} \sqrt{\frac{m-1}{\chi^2_{\alpha/2,m-1}}}, \sigma_{emp} \sqrt{\frac{m-1}{\chi^2_{(1-\alpha)/2,m-1}}}\] 
\vspace{0.05cm}

Since $m>100$, we approximate the chi-square distribution with $m-1$ degrees of freedom by the normal distribution $\mathcal{N}(m-1,2m-2)$. Table~\ref{tab:confidence} provides estimates of the confidence intervals at level $0.001\%$ for different values of $n$, and around $\sigma=3$ and $q/\sqrt{12}$.

\begin{table}[h!]
    \small
    \centering
    \begin{tabular}{rr|cc}
        \toprule
        n & m &  Right ($\sigma=3$)& Wrong ($q/\sqrt{12}$)\\
        \midrule
        80 & 320 &  [2.58, 3.72] & [28.08, 40.45] \\
        150 & 600 &  [2.68, 3.48] & [$1.65 \times 10^3$,$2.15 \times 10^3$]\\
        200 & 800 &  [2.71, 3.40] & [$3.42\times 10^4$, $4.28 \times 10^4$]\\
        256 & 1024 & [2.74, 3.34] & [$1.62 \times 10^6$, $1.98 \times 10^6$]\\
        300 & 1200 & [2.76, 3.31] & [$2.50 \times 10^7$, $3.00 \times 10^7$]\\
        350 & 1400 & [2.78, 3.29] & [$1.02 \times 10^9$, $1.21 \times 10^9$]\\

       \bottomrule
    \end{tabular}
    \vspace{0.1cm}
\caption{\small\textbf{Confidence intervals ($0.001\%$) for secret verification.} Confidence level $0.001\%$. Right: secret is correctly predicted ($\sigma=3$). Wrong: secret is incorrectly predicted ($\sigma=q/\sqrt{12}$). $q$ from Table~\ref{tab:params}. }
    \label{tab:confidence}
    \vspace{-0.6cm}
\end{table}

For instance, for $n=80$, we have $m=320$ and $q=113$. The $0.001\%$ level confidence interval for a correct secret prediction (i.e. measuring $\sigma=3$) is $[2.58, 3.72]$. For an incorrect prediction (measuring $\sigma=q/\sqrt{12}=32.62$), it is $[28.08,40.45]$. Since the two intervals do not overlap, the sample size we use ($m$) is large enough to verify secret guesses (with quasi-certitude). As dimension increases, the confidence intervals grow. This proves our claim that the original $4n$ LWE samples are sufficient to verify model predictions.

\vspace{-0.1cm}
\subsection{Understanding secret recovery}
\label{subsec:more_recoveries}

\begin{figure}[h]

    \vspace{-0.3cm}
    \centering 
    \includegraphics[width = 0.35\textwidth]{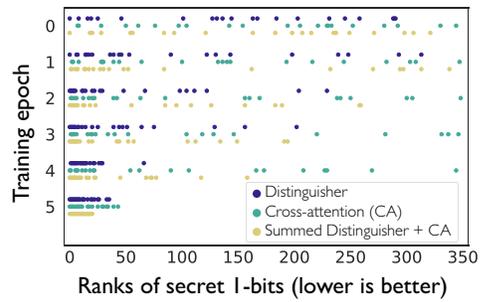}
    \vspace{-0.3cm}
    \caption{\small\textbf{Change in secret bit ranks as training progresses ($n=350$).} }
    \label{fig:learning}
    \vspace{-0.3cm}
\end{figure}

Figure~\ref{fig:learning} shows \system{}'s secret recovery for a successful $n=350$ experiment, in which the combined method recovers the secret in epoch $5$. Figure~\ref{fig:learning} shows how the rankings of the $1$-bits of the secret change throughout training. Our recovery methods guess that the $h$ top-ranked bits are the $1$-bits, so successful recovery occurs when $1$-bits occupy the first $h$ slots on the $x$-axis.

Over time, distinguisher and CA methods learn better ranks for true secret $1$-bits. By epoch $5$, the combined method, which sums the ranks of distinguisher and CA methods, correctly guesses the secret. We do not include direct secret recovery results because it performs poorly for large $n$. Plotting Figure~\ref{fig:learning} requires knowledge of the secret $s$, leveraged here {\it for illustrative purposes only}. \system{} can validate secret guesses without knowledge of $s$, using verification as in \S~\ref{subsec:recovery}.

%% file: tables/table_tinyvslwe.tex
\begin{table}[h!]
\small
  \centering
  \begin{tabular}{cccccccc}
  \toprule
  Setting &
    \multicolumn{7}{c}{{Hamming weight $h$}} \\ \midrule
  \multicolumn{1}{c}{$n=80$} &
    4   & 5   & 6   & 7    & 8    & 9    & 10   \\
  \multicolumn{1}{c}{{TinyLWE}} &
    3/5 & 3/5 & 2/5 & 2/5  & 1/20 & 1/20 & 0/20 \\
  \multicolumn{1}{c}{{LWE}} &
    5/5 & 4/5 & 3/5 & 3/5  & 0/20 & 1/20 & 0/20 \\ \midrule
  \multicolumn{1}{c}{$n=150$} &
    9   & 10  & 11  & 12   & 13   & 14   & 15   \\
  \multicolumn{1}{c}{{\it TinyLWE}} &
    4/5 & 2/5 & 3/5 & 1/5  & 1/20 & 0/20 & 0/20 \\
  \multicolumn{1}{c}{{\it LWE}} &
     5/5 & 2/5 & 2/5 & 1/20 & 0/20 & 0/20 & 0/20 \\ \midrule
  \multicolumn{1}{c}{$n=200$} &
    17  & 18  & 19  & 20   & 21   & 22   & 23   \\
  \multicolumn{1}{c}{{\it TinyLWE}} &
    3/5 & 2/5 & 3/5 & 1/5  & 2/5  & 2/20 & 0/20 \\
  \multicolumn{1}{c}{{\it LWE}} &
  3/5 & 2/5 & 1/5 & 1/5  & 2/20 & 0/20 & 0/20 \\ \midrule
  \multicolumn{1}{c}{$n=256$} &
    26  & 27  & 28  & 29   & 30   & 31   & 32   \\
  \multicolumn{1}{c}{{\it TinyLWE}} &
  4/5 & 1/5 & 1/5 & 3/5  & 3/5  & 4/20 & 0/20 \\
  \multicolumn{1}{c}{\it LWE} &
  3/5 & 2/5 & 3/5 & 3/5  & 1/5  & 3/20 & 2/20 \\ \midrule
  \multicolumn{1}{c}{$n=300$} &
    28  & 29  & 30  & 31   & 32   & 33   & 34   \\ 
  \multicolumn{1}{c}{{\it TinyLWE}} &
    2/5 & 2/5 & 1/5 & 1/5  & 2/5  & 1/5  & 0/20 \\
  \multicolumn{1}{c}{{\it LWE}} &
  1/5 & 3/5 & 2/5 & 1/5  & 1/5  & 0/5  & 0/20 \\ 
  \bottomrule
  \end{tabular}%
  \vspace{0.1cm}
    \caption{\small\textbf{Secret recovery performance: \tinyLWE{} vs. LWE}. Reported values are successes/attempts across different $n$/$h$ settings.}
    \vspace{-0.9cm}
  \label{tab:compare_TinyLWE_lwe}
  \end{table}

%% file: ccs.bbl
\begin{thebibliography}{10}

\bibitem{akkaya2019solving}
{\sc Akkaya, I., Andrychowicz, M., Chociej, M., et~al.}
\newblock Solving rubik's cube with a robot hand, 2019.
\newblock \url{https://arxiv.org/abs/1910.07113}.

\bibitem{HES}
{\sc Albrecht, M., Chase, M., Chen, H., et~al.}
\newblock Homomorphic encryption standard.
\newblock In {\em Protecting Privacy through Homomorphic Encryption}. 2021, pp.~31--62.
\newblock \url{https://eprint.iacr.org/2019/939}.

\bibitem{Albrecht2017_sparse_binary}
{\sc Albrecht, M.~R.}
\newblock {On Dual Lattice Attacks Against Small-Secret LWE and Parameter Choices in HElib and SEAL}.
\newblock In {\em Proc. of EUROCRYPT\/} (2017).

\bibitem{albrecht2017revisiting}
{\sc Albrecht, M.~R., G{\"o}pfert, F., Virdia, F., and Wunderer, T.}
\newblock Revisiting the expected cost of solving usvp and applications to lwe.
\newblock In {\em {Proc. of ASIACRYPT}\/} (2017).

\bibitem{LWEestimator}
{\sc Albrecht, M.~R., Player, R., and Scott, S.}
\newblock On the concrete hardness of learning with errors.
\newblock {\em Journal of Mathematical Cryptology 9}, 3 (2015), 169--203.

\bibitem{crys_kyber}
{\sc Avanzi, R., Bos, J., Ducas, L., Kiltz, E., Lepoint, T., Lyubashevsky, V., Schanck, J.~M., Schwabe, P., Seiler, G., and Stehlé, D.}
\newblock {CRYSTALS-Kyber (version 3.02) – Submission to round 3 of the NIST post-quantum project.}
\newblock Available at \url{https://pq-crystals.org/}.

\bibitem{bahdanau2014}
{\sc Bahdanau, D., Cho, K., and Bengio, Y.}
\newblock Neural machine translation by jointly learning to align and translate.
\newblock In {\em Proc. of ICLR\/} (2014).

\bibitem{bai2019refined}
{\sc Bai, S., Miller, S., and Wen, W.}
\newblock {A refined analysis of the cost for solving LWE via uSVP}.
\newblock In {\em {Proc. of ASIACRYPT}\/} (2019).

\bibitem{BLPRS13_poly_modulus_hardness}
{\sc Brakerski, Z., Langlois, A., Peikert, C., Regev, O., and Stehl\'{e}, D.}
\newblock {Classical Hardness of Learning with Errors}.
\newblock In {\em Proc. of the ACM Symposium on Theory of Computing\/} (2013).

\bibitem{carion2020endtoend}
{\sc Carion, N., Massa, F., Synnaeve, G., et~al.}
\newblock End-to-end object detection with transformers, 2020.
\newblock \url{https://arxiv.org/abs/2005.12872}.

\bibitem{charton2021linear}
{\sc Charton, F.}
\newblock Linear algebra with transformers, 2021.
\newblock \url{https://arxiv.org/abs/2112.01898}.

\bibitem{CCLS}
{\sc Chen, H., Chua, L., Lauter, K., and Song, Y.}
\newblock {On the Concrete Security of LWE with Small Secret}.
\newblock Cryptology ePrint Archive, Paper 2020/539, 2020.
\newblock \url{https://eprint.iacr.org/2020/539}.

\bibitem{nist2022finalists}
{\sc Chen, L., Moody, D., Liu, Y.-K., et~al.}
\newblock {PQC Standardization Process: Announcing Four Candidates to be Standardized, Plus Fourth Round Candidates}.
\newblock {\em US Department of Commerce, NIST\/} (2022).
\newblock \url{https://csrc.nist.gov/News/2022/pqc-candidates-to-be-standardized-and-round-4}.

\bibitem{CN11_BKZ}
{\sc Chen, Y., and Nguyen, P.~Q.}
\newblock {BKZ 2.0: Better Lattice Security Estimates}.
\newblock In {\em Proc. of ASIACRYPT\/} (2011).

\bibitem{Cheon_hybrid_dual}
{\sc Cheon, J.~H., Hhan, M., Hong, S., and Son, Y.}
\newblock {A Hybrid of Dual and Meet-in-the-Middle Attack on Sparse and Ternary Secret LWE}.
\newblock {\em IEEE Access\/} (2019).

\bibitem{heaan}
{\sc Cheon, J.~H., Kim, A., Kim, M., and Song, Y.}
\newblock Homomorphic encryption for arithmetic of approximate numbers.
\newblock In {\em Proc. of ASIACRYPT\/} (2017).

\bibitem{Lizard}
{\sc Cheon, J.~H., Kim, D., Lee, J., and Song, Y.}
\newblock {Lizard: Cut Off the Tail! A Practical Post-quantum Public-Key Encryption from LWE and LWR}.
\newblock In {\em Security and Cryptography for Networks\/} (2018).

\bibitem{seq2seq2014}
{\sc Cho, K., van Merrienboer, B., Gulcehre, C., et~al.}
\newblock Learning phrase representations using rnn encoder-decoder for statistical machine translation.
\newblock In {\em Proc. of EMNLP\/} (2014).

\bibitem{csordas2021neural}
{\sc Csordás, R., Irie, K., and Schmidhuber, J.}
\newblock {The Neural Data Router: Adaptive Control Flow in Transformers Improves Systematic Generalization}.
\newblock In {\em {Proc. of ICML}\/} (2022).

\bibitem{Rachel_Player_sparse}
{\sc Curtis, B.~R., and Player, R.}
\newblock On the feasibility and impact of standardising sparse-secret {LWE} parameter sets for homomorphic encryption.
\newblock In {\em Proc. of the {ACM} Workshop on Encrypted Computing {\&} Applied Homomorphic Cryptography\/} (2019).

\bibitem{dehghani2018universal}
{\sc Dehghani, M., Gouws, S., Vinyals, O., Uszkoreit, J., and Kaiser, {\L}.}
\newblock Universal transformers.
\newblock In {\em {Proc. of ICLR}\/} (2019).

\bibitem{fplll}
{\sc development team, T.~F.}
\newblock {fplll}, a lattice reduction library, {Version}: 5.4.4.
\newblock Available at \url{https://github.com/fplll/fplll}, 2023.

\bibitem{DongSpeechTransformer}
{\sc Dong, L., Xu, S., and Xu, B.}
\newblock Speech-transformer: A no-recurrence sequence-to-sequence model for speech recognition.
\newblock In {\em Proc. of ICASSP\/} (2018).

\bibitem{dosovitskiy2021image}
{\sc Dosovitskiy, A., Beyer, L., Kolesnikov, A., et~al.}
\newblock An image is worth 16x16 words: Transformers for image recognition at scale.
\newblock In {\em Proc. of ICLR\/} (2021).

\bibitem{crys_dilithium}
{\sc Ducas, L., Kiltz, E., Lepoint, T., Lyubashevsky, V., Schwabe, P., Seiler, G., and Stehlé, D.}
\newblock {CRYSTALS-Dilithium – Algorithm Specifications and Supporting Documentation (Version 3.1)}.
\newblock Available at \url{https://pq-crystals.org/}.

\bibitem{Kan87}
{\sc Kannan, R.}
\newblock {Minkowski's Convex Body Theorem and Integer Programming}.
\newblock {\em Mathematics of Operations Research 12\/} (1987), 415--440.

\bibitem{kingma2014adam}
{\sc Kingma, D.~P., and Ba, J.}
\newblock Adam: A method for stochastic optimization.
\newblock In {\em {Proc. of ICLR}\/} (2015).

\bibitem{laine2015key}
{\sc Laine, K., and Lauter, K.}
\newblock Key recovery for lwe in polynomial time.
\newblock {\em Cryptology ePrint Archive\/} (2015).
\newblock \url{https://eprint.iacr.org/2015/176.pdf}.

\bibitem{lample2019deep}
{\sc Lample, G., and Charton, F.}
\newblock Deep learning for symbolic mathematics.
\newblock In {\em Proc. of ICLR\/} (2020).

\bibitem{LLL}
{\sc Lenstra, H.~j., Lenstra, A., and Lovász, L.}
\newblock Factoring polynomials with rational coefficients.
\newblock {\em Mathematische Annalen 261\/} (1982), 515--534.

\bibitem{li2023salsa}
{\sc Li, C., Sotakova, J., Wenger, E., Allen-Zhu, Z., Charton, F., and Lauter, K.}
\newblock Salsa verde: a machine learning attack on learning with errors with sparse small secrets, 2023.
\newblock \url{https://arxiv.org/abs/2306.11641}.

\bibitem{LM09_hardness_lwe_exp}
{\sc Lyubashevsky, V., and Micciancio, D.}
\newblock On bounded distance decoding, unique shortest vectors, and the minimum distance problem.
\newblock In {\em Proc. of CRYPTO\/} (2009), S.~Halevi, Ed.

\bibitem{MV_SVP_exp}
{\sc Micciancio, D., and Voulgaris, P.}
\newblock Faster exponential time algorithms for the shortest vector problem.
\newblock In {\em Proceedings of the ACM-SIAM Symposium on Discrete Algorithms\/} (2010).

\bibitem{palamasinvestigating}
{\sc Palamas, T.}
\newblock Investigating the ability of neural networks to learn simple modular arithmetic.

\bibitem{Pei09a}
{\sc Peikert, C.}
\newblock {Public-Key Cryptosystems from the Worst-Case Shortest Vector Problem: Extended Abstract}.
\newblock In {\em Proc. of the ACM Symposium on Theory of Computing\/} (2009).

\bibitem{polu2020generative}
{\sc Polu, S., and Sutskever, I.}
\newblock Generative language modeling for automated theorem proving, 2020.
\newblock \url{https://arxiv.org/abs/2009.03393}.

\bibitem{radford2018improving}
{\sc Radford, A., Narasimhan, K., Salimans, T., and Sutskever, I.}
\newblock Improving language understanding by generative pre-training.
\newblock {\em OpenAI blog\/} (2018).
\newblock \url{https://s3-us-west-2.amazonaws.com/openai-assets/research-covers/language-unsupervised/language_understanding_paper.pdf}.

\bibitem{radford2019language}
{\sc Radford, A., Wu, J., Child, R., Luan, D., Amodei, D., Sutskever, I., et~al.}
\newblock Language models are unsupervised multitask learners.
\newblock {\em OpenAI blog\/} (2019).
\newblock \url{https://d4mucfpksywv.cloudfront.net/better-language-models/language_models_are_unsupervised_multitask_learners.pdf}.

\bibitem{dalle2021}
{\sc Ramesh, A., Pavlov, M., Goh, G., Gray, S., Voss, C., Radford, A., Chen, M., and Sutskever, I.}
\newblock Zero-shot text-to-image generation, 2021.
\newblock \url{https://arxiv.org/abs/2102.12092}.

\bibitem{regev:quantum}
{\sc Regev, O.}
\newblock Quantum computation and lattice problems.
\newblock {\em SIAM Journal on Computing 33}, 3 (2004), 738--760.

\bibitem{Reg05}
{\sc Regev, O.}
\newblock {On Lattices, Learning with Errors, Random Linear Codes, and Cryptography}.
\newblock In {\em Proc. of the ACM Symposium on Theory of Computing\/} (2005).

\bibitem{rivest1978method}
{\sc Rivest, R.~L., Shamir, A., and Adleman, L.}
\newblock A method for obtaining digital signatures and public-key cryptosystems.
\newblock {\em Communications of the ACM\/} (1978).

\bibitem{rogez2017lcr}
{\sc Rogez, G., Weinzaepfel, P., and Schmid, C.}
\newblock Lcr-net: Localization-classification-regression for human pose.
\newblock In {\em Proc. of CVPR\/} (2017).

\bibitem{rothe2015dex}
{\sc Rothe, R., Timofte, R., and Van~Gool, L.}
\newblock Dex: Deep expectation of apparent age from a single image.
\newblock In {\em Proc. of ICCV\/} (2015).

\bibitem{BKZ}
{\sc Schnorr, C.}
\newblock A hierarchy of polynomial time lattice basis reduction algorithms.
\newblock {\em Theoretical Computer Science 53}, 2 (1987), 201--224.

\bibitem{schnorr_euchner}
{\sc Schnorr, C.~P., and Euchner, M.}
\newblock Lattice basis reduction: Improved practical algorithms and solving subset sum problems.
\newblock {\em Mathematical Programming 66}, 1-3 (Aug. 1994), 181--199.

\bibitem{schrittwieser2020mastering}
{\sc Schrittwieser, J., Antonoglou, I., Hubert, T., et~al.}
\newblock {Mastering ATARI, Go, Chess and Shogi by planning with a learned model}.
\newblock {\em Nature 588\/} (2020), 604--609.

\bibitem{SEAL}
{M}icrosoft {SEAL} (release 4.1).
\newblock \url{https://github.com/Microsoft/SEAL}, Jan. 2023.
\newblock Microsoft Research, Redmond, WA.

\bibitem{transformer17}
{\sc Vaswani, A., Shazeer, N., Parmar, N., et~al.}
\newblock Attention is all you need.
\newblock In {\em Proc. of NeurIPS\/} (2017).

\bibitem{Wang_2020}
{\sc Wang, Y., Mohamed, A., Le, D., et~al.}
\newblock Transformer-based acoustic modeling for hybrid speech recognition.
\newblock {\em Proc. of ICASSP\/} (2020).

\bibitem{wengersalsa}
{\sc Wenger, E., Chen, M., Charton, F., and Lauter, K.}
\newblock Salsa: Attacking lattice cryptography with transformers.
\newblock In {\em Proc. of NeurIPS\/} (2022).

\end{thebibliography}
